\documentstyle[psfig]{mn}
\newif\ifAMStwofonts
\newcommand{\lapp}{\mbox{\raisebox{-0.3em}{$\stackrel{\textstyle <}{\sim}$}}}
\newcommand{\gapp}{\mbox{\raisebox{-0.3em}{$\stackrel{\textstyle >}{\sim}$}}}

\title{Radio observations of a few giant sources}
\author[C. Konar et al.]
       {C. Konar$^{1}$$\thanks{E-mail: sckonar@ncra.tifr.res.in}$, D.J. Saikia$^{1,2}$$\thanks{E-mail: djs@ncra.tifr.res.in, djs@jb.man.ac.uk}$, 
C.H. Ishwara-Chandra$^{1}$ and V.K. Kulkarni$^{1}$ \\
$^{1}$ National Centre for Radio Astrophysics, TIFR, Pune University Campus, Post Bag 3,
Pune 411 007, India \\
$^{2}$ Jodrell Bank Observatory, University of Manchester, Macclesfield, Cheshire, SK11 9DL, UK \\
}
\date{Accepted.    Received }

\pagerange{\pageref{firstpage}--\pageref{lastpage}}
\pubyear{  }

\begin{document}

\maketitle

\label{firstpage}

\begin{abstract}
We present multifrequency observations with the 
Giant Metrewave Radio Telescope (GMRT) and the Very Large Array (VLA) 
of a sample of seventeen largely giant radio sources (GRSs). These  
observations have either helped clarify the radio structures or provided new
information at a different frequency.  The broad line
radio galaxy, J0313+413, has an asymmetric, curved radio jet and a variable radio
core, consistent with a moderate angle of inclination to the line of sight. We attempt
to identify steep spectrum radio cores (SSCs), which may be a sign of recurrent activity,
and find four candidates. If confirmed, this
would indicate a trend for SSCs to occur preferentially in GRSs. From the 
structure and integrated spectra of the sources, we suggest that the lobes of emission
in J0139+399 and J0200+408 may be due to an earlier cycle of nuclear activity.
We find that inverse-Compton losses with the cosmic microwave background radiation dominate
over synchrotron radiative losses in the lobes of all the sources, consistent
with earlier studies. We also show that the prominence of the bridge emission 
decreases with increasing redshift, possibly due to inverse-Compton losses. This could affect
the appearance and identification of GRSs at large redshifts. 
\end{abstract}


\section{Introduction}
Giant radio sources (GRSs), defined to be those with a projected linear size
$\gapp$1 Mpc (H$_o$=71 km s$^{-1}$ Mpc$^{-1}$, $\Omega_m$=0.27,
$\Omega_{vac}$=0.73, Spergel et al. 2003) are useful for studying the late stages of the  
evolution of radio sources, constraining orientation-dependent unified schemes
and probing the intergalactic medium at different redshifts 
(e.g. Subrahmanyan \& Saripalli 1993; Subrahmanyan, Saripalli \& Hunstead 1996;
Mack et al. 1998; Ishwara-Chandra \& Saikia 1999, 
hereinafter referred to as IC99; Kaiser \& Alexander 1999; Blundell, Rawlings \& Willott 1999 and 
references therein; Schoenmakers 1999, hereinafter referred to as S99;  Schoenmakers et al. 2000a, 2001). 
Recent attempts to identify GRSs from large radio surveys such as WENSS
(Westerbork Northern Sky Survey), NVSS (NRAO VLA Sky Survey) and
FIRST (Faint Images of the Radio Sky at Twenty-cm) have significantly increased
their numbers (S99; Machalski, Jamrozy \& Zola 2001; Chy\.{z}y et al. 2003), 
but there is still a dearth of giants
$\gapp$2 Mpc and at cosmologically interesting redshifts of $\gapp$1. 
The highest redshift giant source known so
far is 4C 39.24, which is associated with a galaxy at a redshift of 1.88 and has been
studied in some detail by Law-Green et al. (1995), while the most distant
giant quasar is J1432+158 (Singal, Konar \& Saikia 2004) at a redshift of 1.005.

In this paper we first report GMRT and VLA observations of a sample of largely giant sources.
A few of the sources were earlier classified as GRSs, but now have sizes smaller 
than 1 Mpc, based on a revised value of the Hubble constant. We have nevertheless included
these sources in this paper, since there is a continuity of sizes and the limit of a Mpc is a 
working definition to identify those in the late stages of evolution.
These observations have either helped clarify the radio structures or provided new
information on these sources at a different frequency.  We examine 
the radiative losses in these sources, and show that inverse-Compton losses dominate
over synchrotron radiative losses in the lobes, consistent with earlier studies (e.g. IC99; 
Schoenmakers et al. 2000a).
In order to develop strategies for indentifying GRSs at high redshifts we investigate
the prominence of bridge emission, f$_{\rm bridge}$, as a function of redshift, and find that f$_{\rm bridge}$
decreases with increasing redshift. These diffuse regions of emission
are more likely to be affected by inverse-Compton losses, which would affect the appearance and
identification of giants at large redshifts. We explore the evidence for the steep-spectrum
cores or SSCs ($\alpha_{\rm core}\lapp-$0.5, S$\propto\nu^{\alpha}$), which might indicate 
renewed activity, being more common in GRSs compared with
the smaller-sized objects, and identify candidate SSCs in our sample. Assuming that nuclear activity may be
recurrent, we attempt to identify lobes from earlier episodes of nuclear activity on the basis of the structure and
their overall radio spectra.

\begin{table}
\caption{ Observing log }
\begin{tabular}{l l c l c }

\hline
Teles-    & Array  & Obs.   &  Sources  & Obs.  \\
cope      & Conf.  & Freq.  &           & Date              \\
          &        & GHz    &           &                \\
\hline

GMRT      &        & 0.6    & J0657+481            & 2003Jan10 \\

GMRT      &        & 0.6    & J0313+413, J1313+695 & 2003Sep06 \\

GMRT      &        & 0.6    & J1604+375, J2042+751 & 2004Jan01  \\

GMRT      &        & 0.6    & J1702+422            & 2004Jan17  \\

GMRT      &        & 1.3    & J0139+399            & 2003Aug19 \\

VLA       &  BnC   & 1.4    & J0657+481, J1101+365 & 2000Mar10 \\
          &        &        & J1200+348, J1235+213 &           \\
          &        &        & J1313+695            &           \\

VLA       &  BnC   & 1.4    & J0135+379, J0313+413 & 2000Mar13 \\
          &        &        & J1604+375, J1637+417 &           \\
          &        &        & J1702+422, J2312+187 &           \\

VLA       &  D     & 5      & J0135+379, J0139+399 & 2000Jul24 \\
          &        &        & J0200+408, J0754+432 &           \\
          &        &        & J0819+756, J1101+365 &           \\
          &        &        & J1919+517, J2042+751 &           \\
          &        &        & J2312+187            &           \\

VLA       &  D     & 5      & J1200+348, J1313+696 & 2000Aug18 \\
          &        &        & J1604+375, J1637+417 &           \\
          &        &        & J1702+422            &           \\

VLA       &  D     & 5      & J0657+481            & 2000Sep11 \\

\hline
\end{tabular}
\end{table}

\section{Observations and analyses} 
The GMRT consists of thirty 45-m antennas in an approximate `Y'
shape similar to the VLA but with each antenna in a fixed position.
Twelve antennas are randomly placed within a central
1 km by 1 km square (the ``Central Square'') and the
remainder form the irregularly shaped Y (6 on each arm) over a total extent of about 25 km.
Further details about the array can be found at the GMRT website
at {\tt http://www.gmrt.ncra.tifr.res.in}.
The observations were made in the standard fashion, with
each source observation interspersed with observations
of the phase calibrator. The primary flux density calibrator was
either 3C48, 3C147 or 3C286, with all flux densities being on the
Baars et al. (1977) scale. The observing time on the source varied from about an
hour to a few hours. However, the low-frequency data were sometimes
significantly affected by ionospheric disturbances.

\begin{figure}
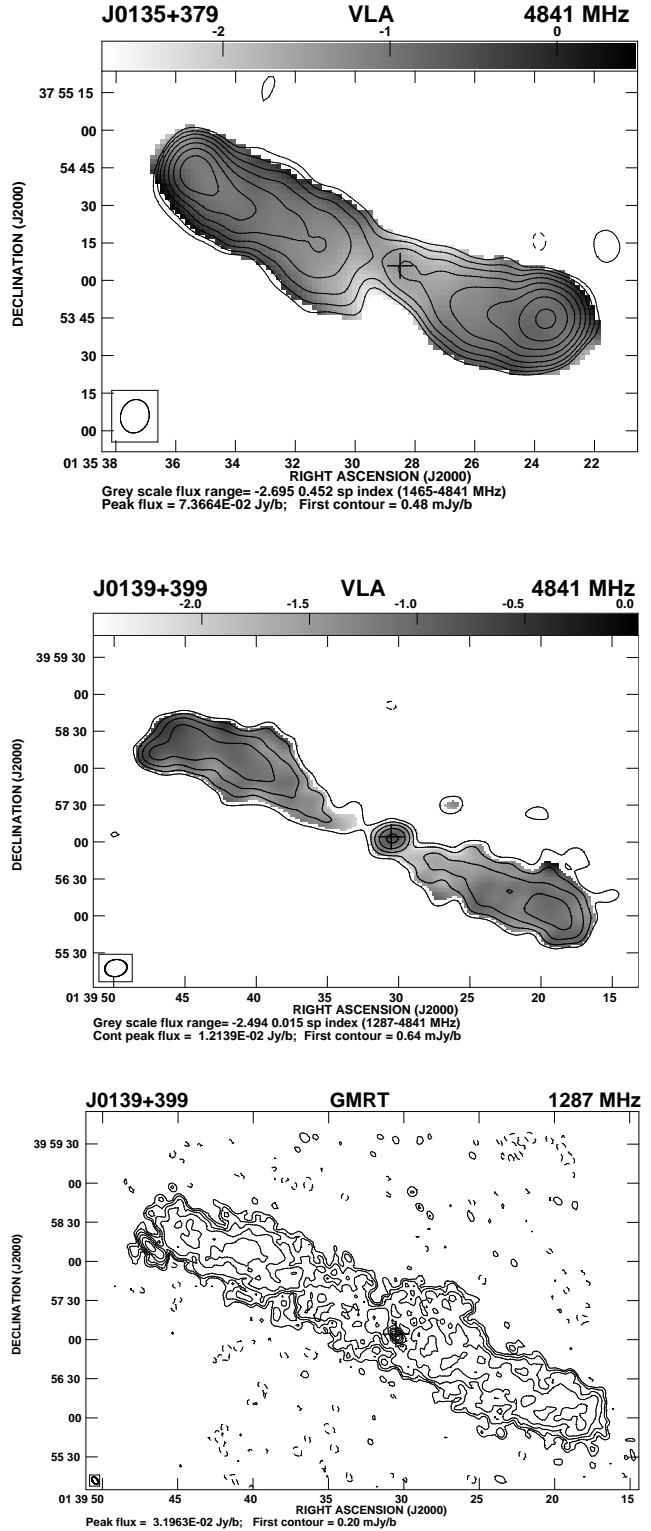

\vbox{
  \psfig{file=J0135+379.C+SPIX.ps,width=3.35in,angle=-90}
   \psfig{file=J0139+399.C+SPIX.ps,width=3.35in,angle=-90}
   \psfig{file=J0139+399L.GMT.ps,width=3.35in,angle=-90}
    }
\caption[Radio images of the sources]{Radio images of the sources. The contour levels for all the images 
are -1, 1, 2, 4, 8, 16, \ldots times the first contour level. The peak brightness in the image in units of Jy/beam 
and the level of 
the first contour in units of mJy/beam are given below the images. The restoring beam is indicated by an ellipse.
The $+$ sign indicates the positions of the optical identification 
available in the literature. In four of the images the 
spectral index distribution
in gray is superimposed on the total-intensity contours.
}
\end{figure}

The observations with the VLA were made in the snap-shot mode in the 
L and C bands with approximately 10 to 20 minutes on the source. The flux
densities are again on the Baars et al. (1977) scale.  All the data were
calibrated and analysed in the standard way using the NRAO {\tt AIPS} 
package. Spectral indices were estimated from our data only for regions or 
features which were unlikely to be significantly affected by any missing 
flux density. 

The observing log for both the GMRT and the VLA observations is given in 
Table 1 which is arranged as follows. Columns 1 and 2 show the name of the 
telescope, and the array configuration for the VLA observations;
column 3 shows the approximate frequency of the observations in GHz, while the sources observed and
the dates of the observations are listed in Columns 4 and 5 respectively. The precise
frequencies in MHz are shown in each image and are also listed in Table 2.

\begin{figure}
\vbox{
    \psfig{file=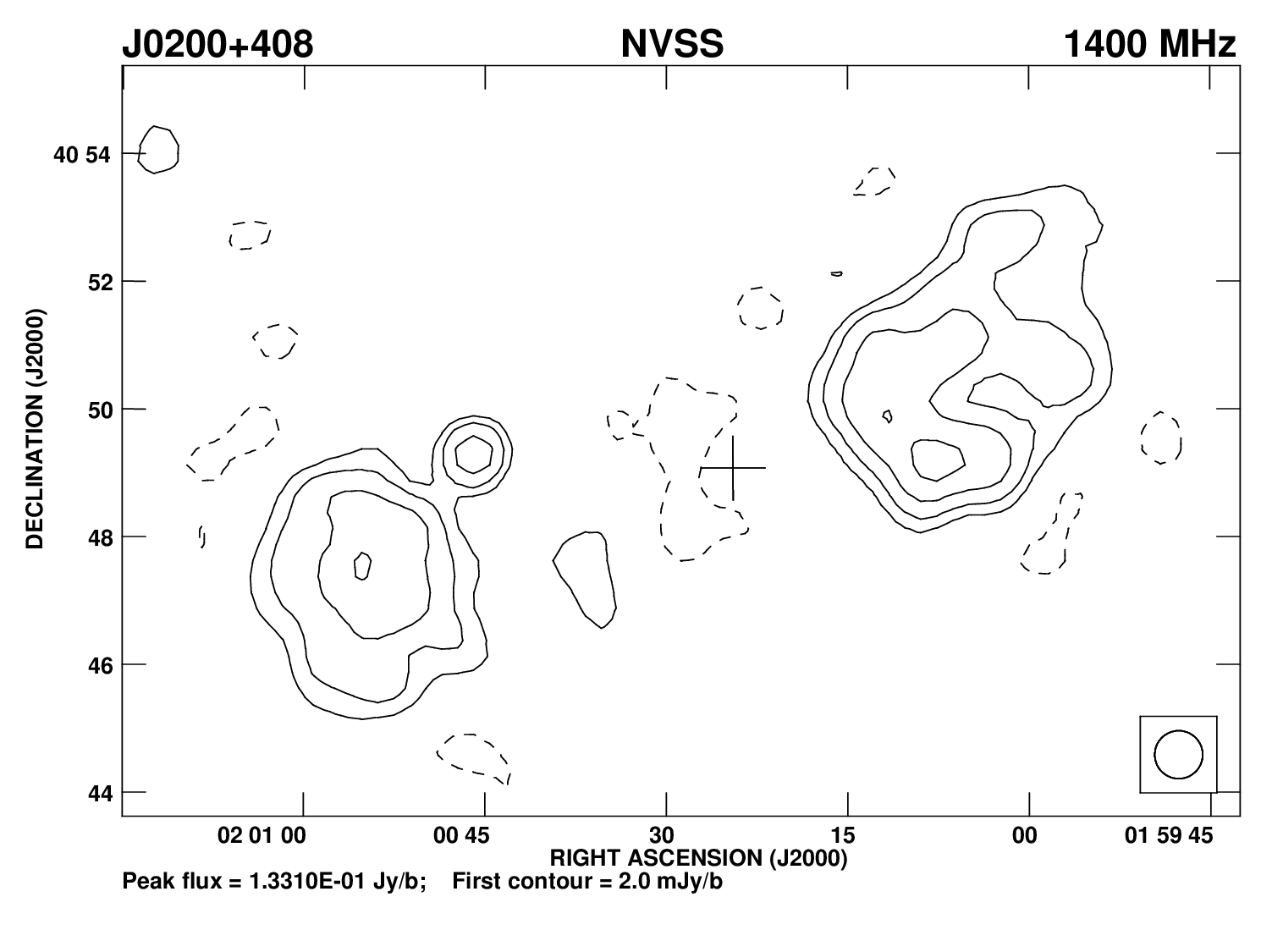,width=3.25in,angle=-90}
    \psfig{file=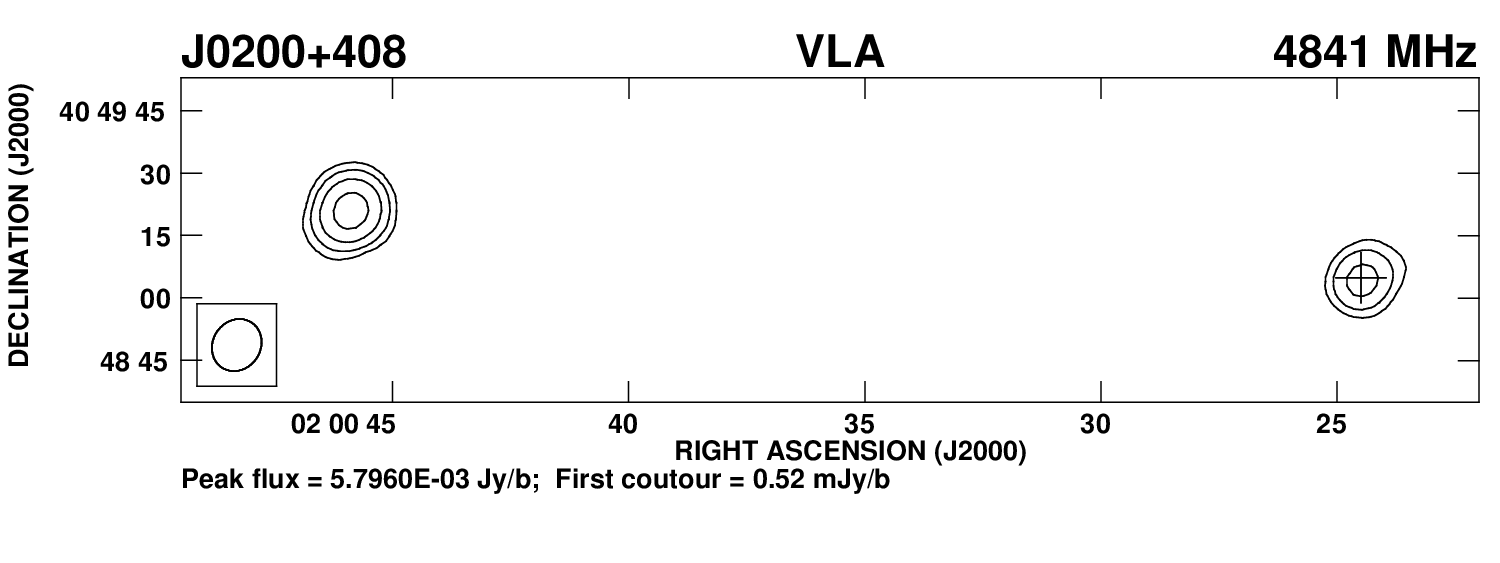,width=3.25in,angle=-90}
    }
\contcaption{}
\end{figure}

\begin{figure}
\vbox{
    \psfig{file=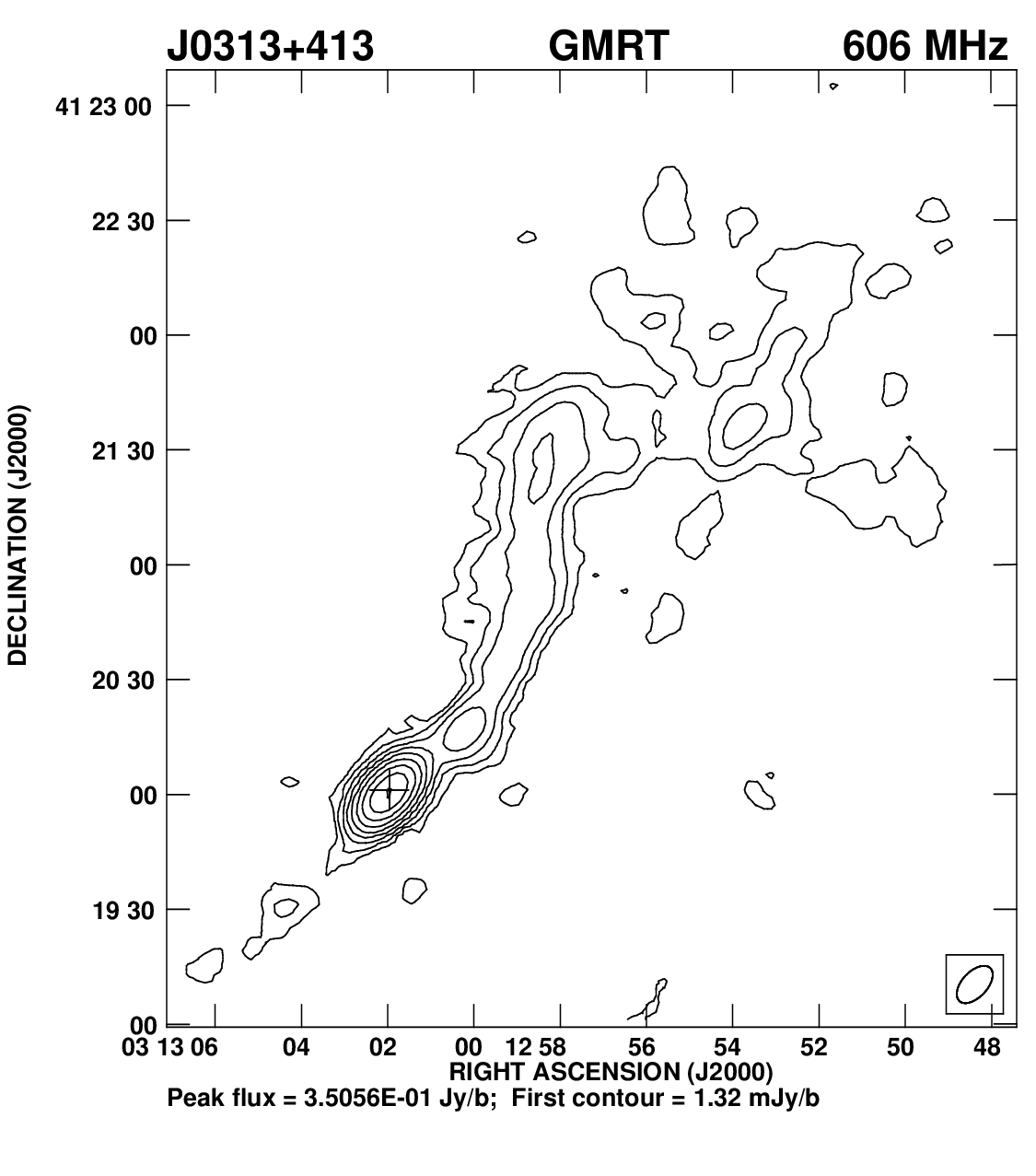,width=3.25in,angle=0}
    \psfig{file=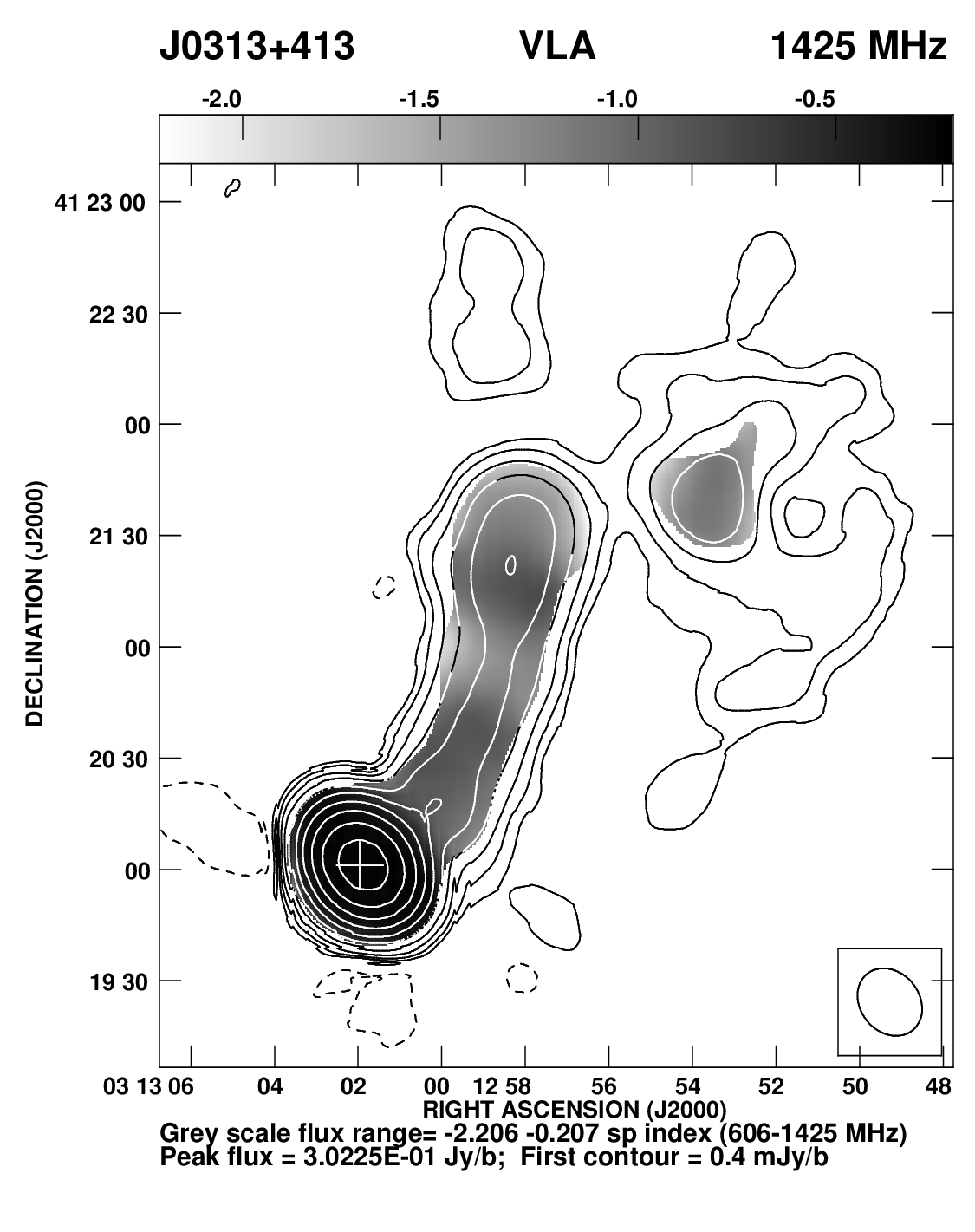,width=3.25in,angle=0}
    }
\contcaption{}
\end{figure}

\section{Observational results}

The images of the sources are presented in
Figure 1, while the observational parameters and some of the observed
properties are presented in Table 2, which is arranged as follows.
Column 1: Name of the source; column 2: frequency  of observations in units of MHz, and
the letter G or V representing either GMRT or VLA observations;
columns 3-5: the major and minor axes of the restoring beam in arcsec and its PA in degrees;
column 6: the rms noise in units of mJy/beam; column 7: the integrated flux density of the
source in mJy estimated by specifying an area enclosing the entire source. We examined the
change in flux density by specifying different areas and found the difference to be within
a few per cent. The flux densities at different frequencies have been estimated over 
similar areas.  
Columns 8, 11 and 14: component designation, where W, E, N, S and C denote the western, eastern,
northern, southern and core components respectively;
columns 9 and 10, 12 and 13, and 15 and 16: the peak and total flux densities of each of the
components in units of mJy/beam and mJy. The core flux densities were sometimes evaluated by imaging the
source using the longer spacings so that the core appears reasonably isolated. The superscript $g$ 
indicates that the flux densities have been estimated from a two-dimensional Gaussian fit to the core component.
A $?$ denotes a possible detection of a core component which requires further confirmation.

\begin{figure*}
\hbox{
% \hspace{-0.5cm}
  \vbox{
  \psfig{file=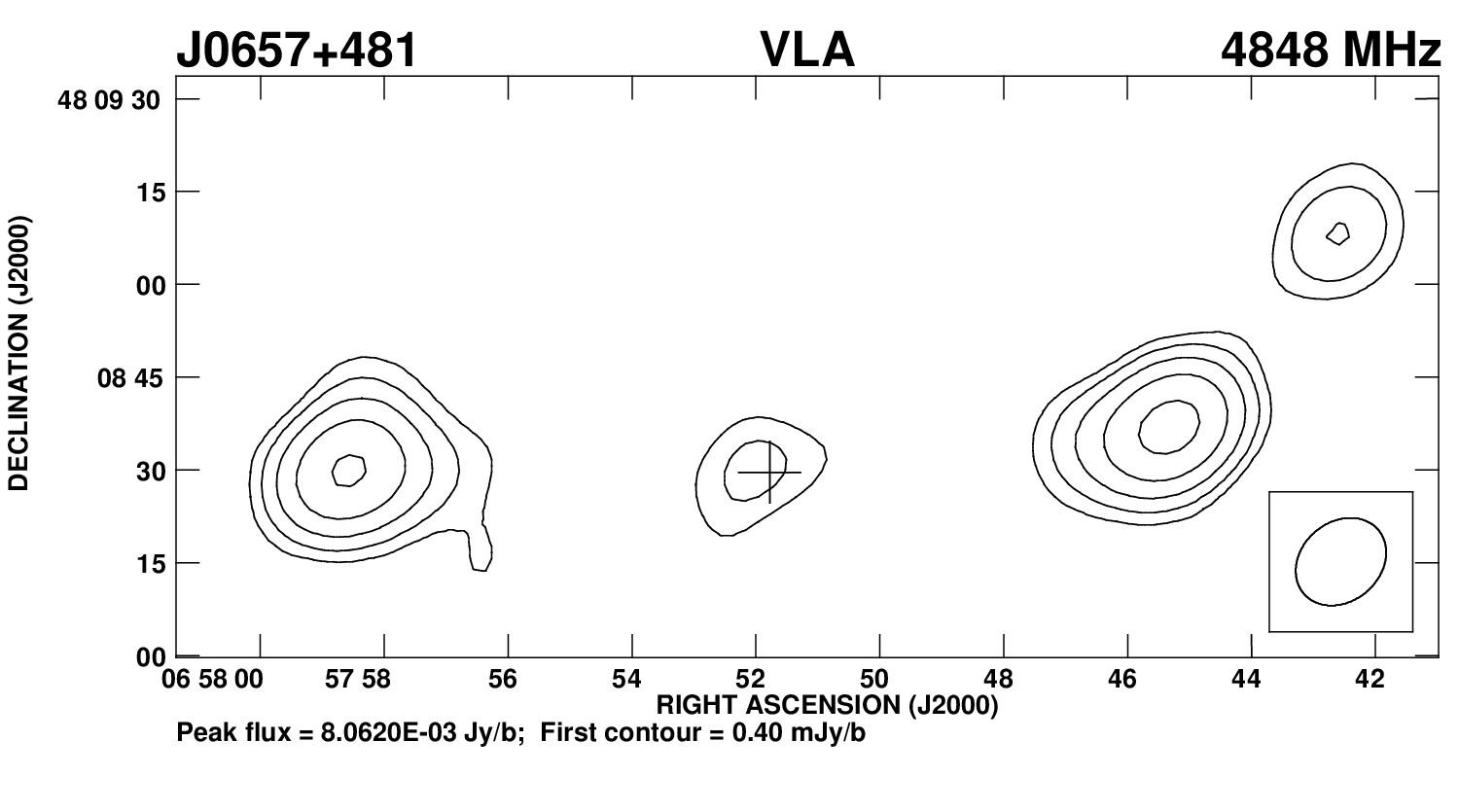,width=2.5in,angle=-90}
  \psfig{file=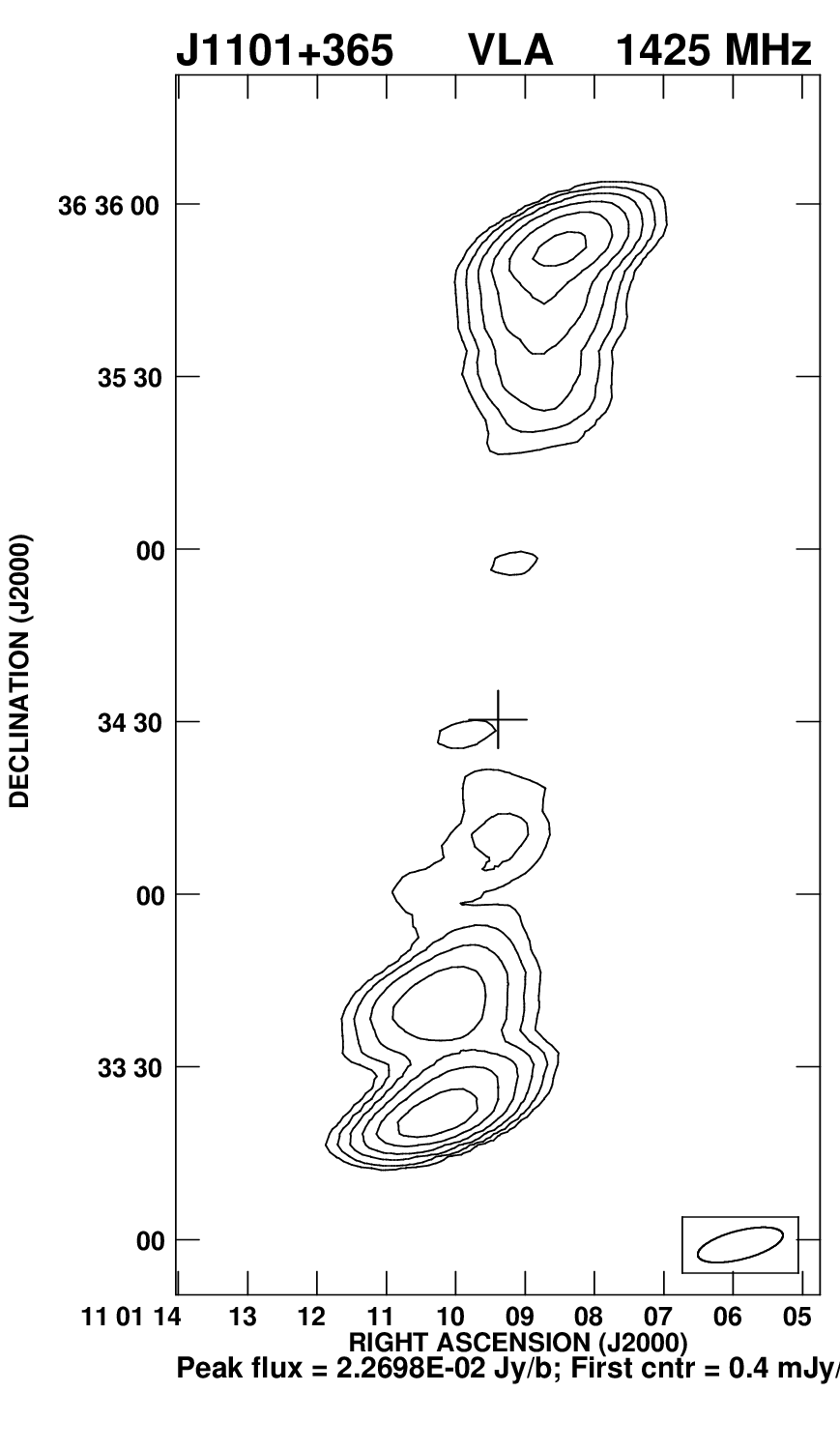,width=1.8in,angle=0}
  \psfig{file=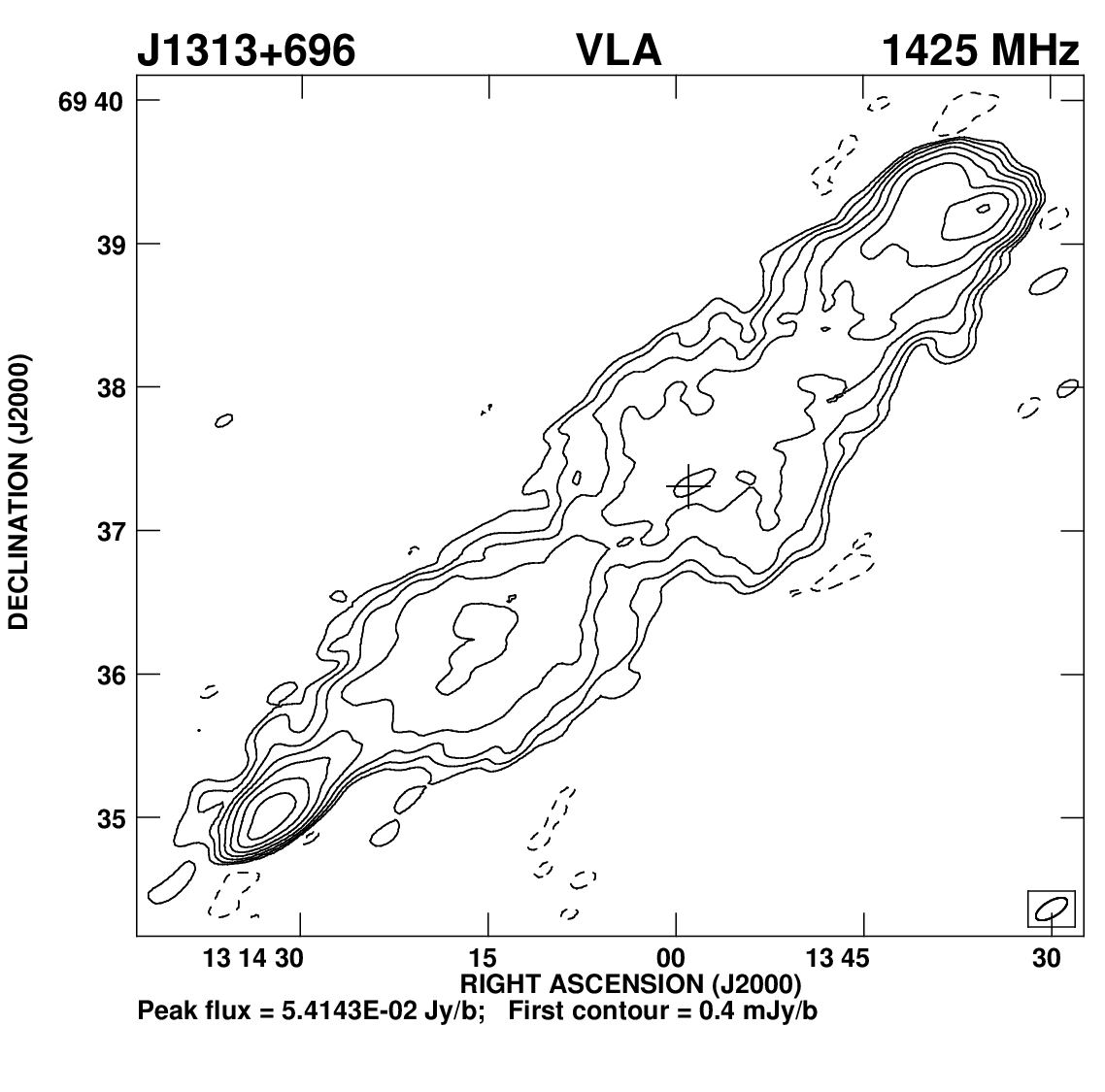,width=2.5in,angle=-90}
  }
  \vbox{
% \vspace{-0.5cm}
  \psfig{file=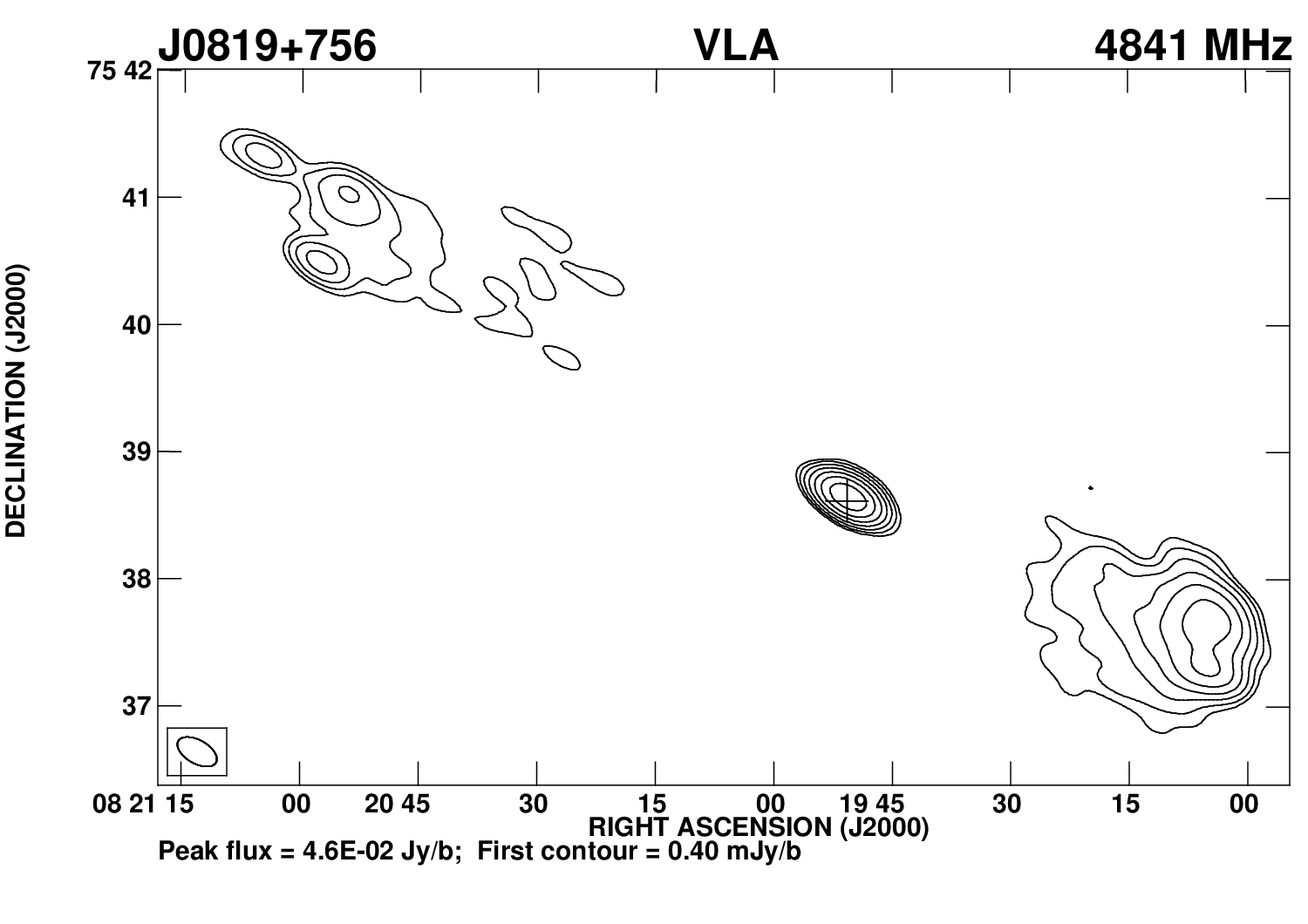,width=2.5in,angle=-90}
  \psfig{file=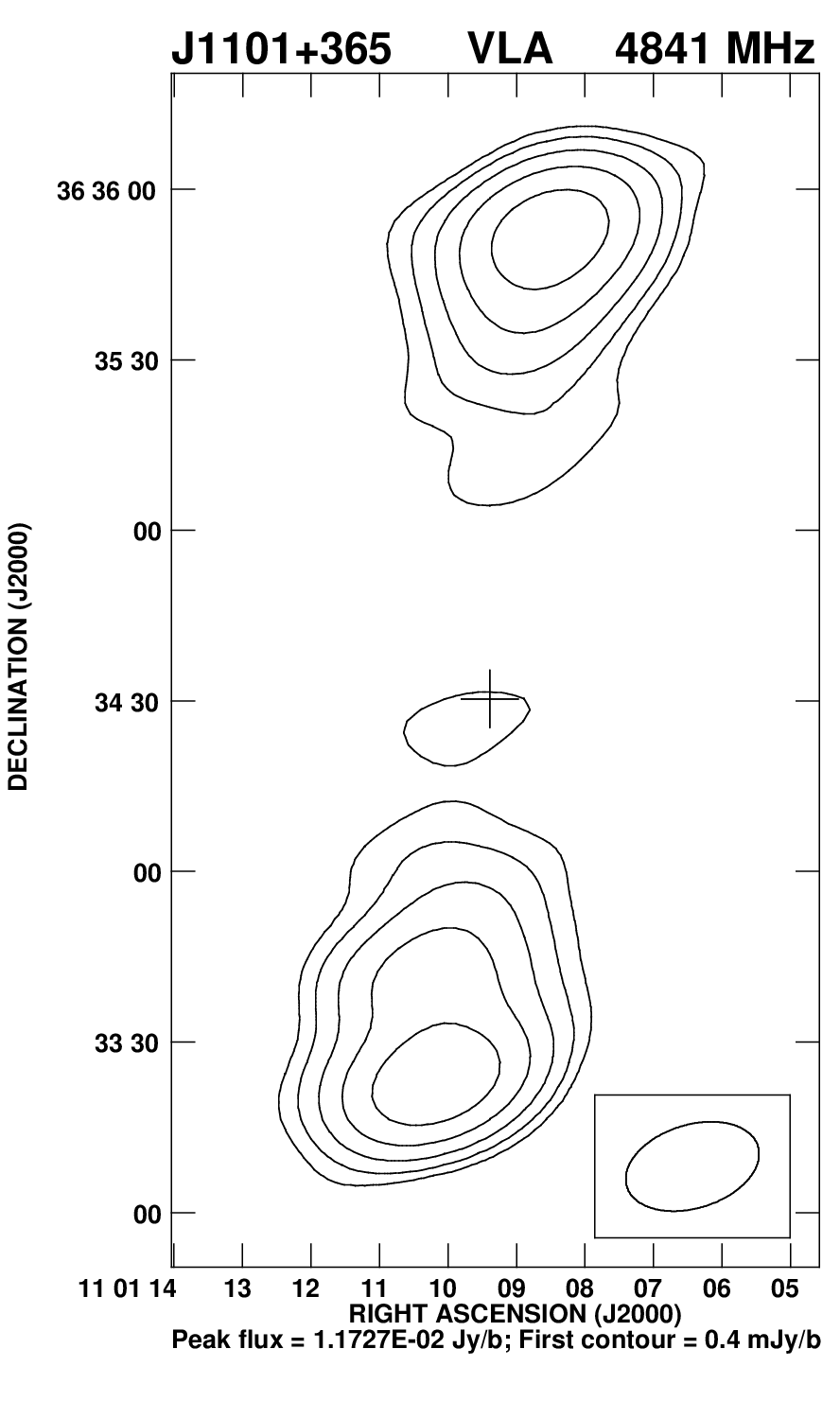,width=1.8in,angle=0}
  \psfig{file=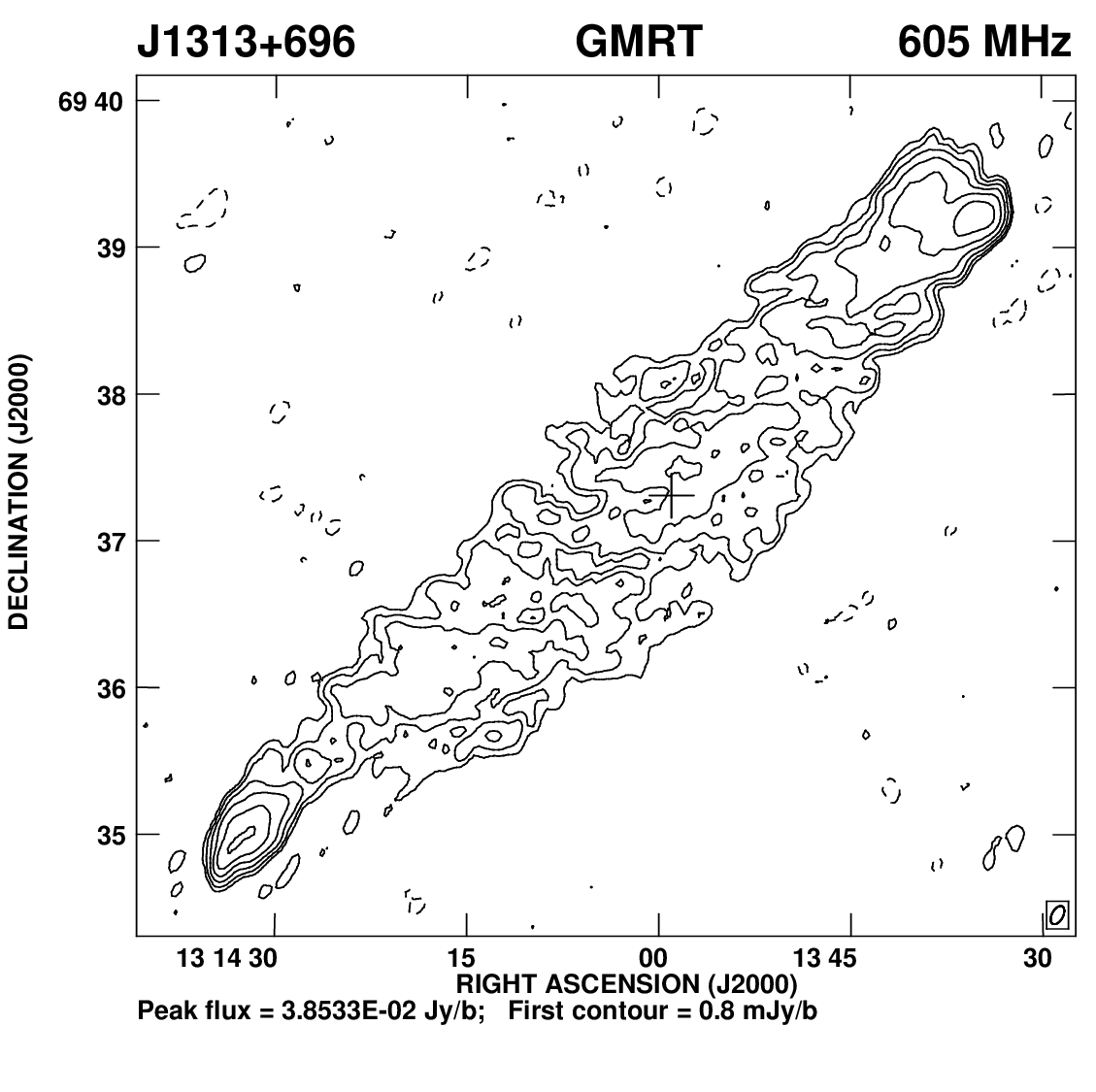,width=2.5in,angle=-90}
  }
  \vbox{
  \psfig{file=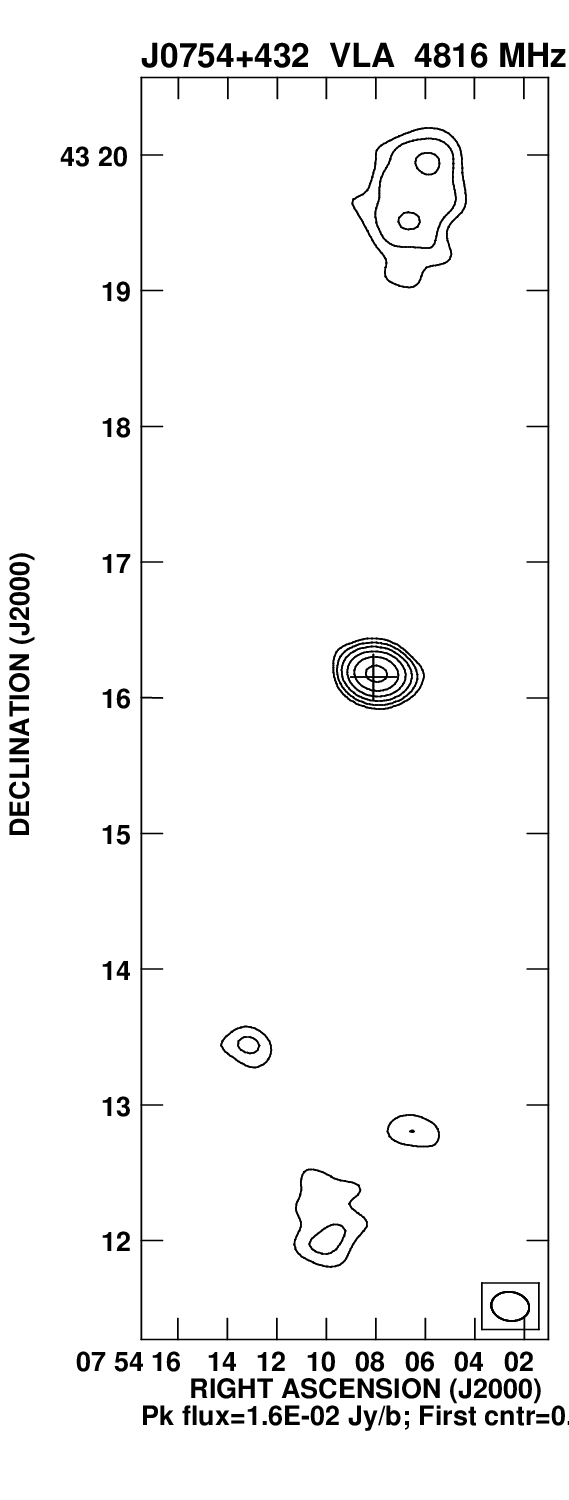,width=1.8in}
  \psfig{file=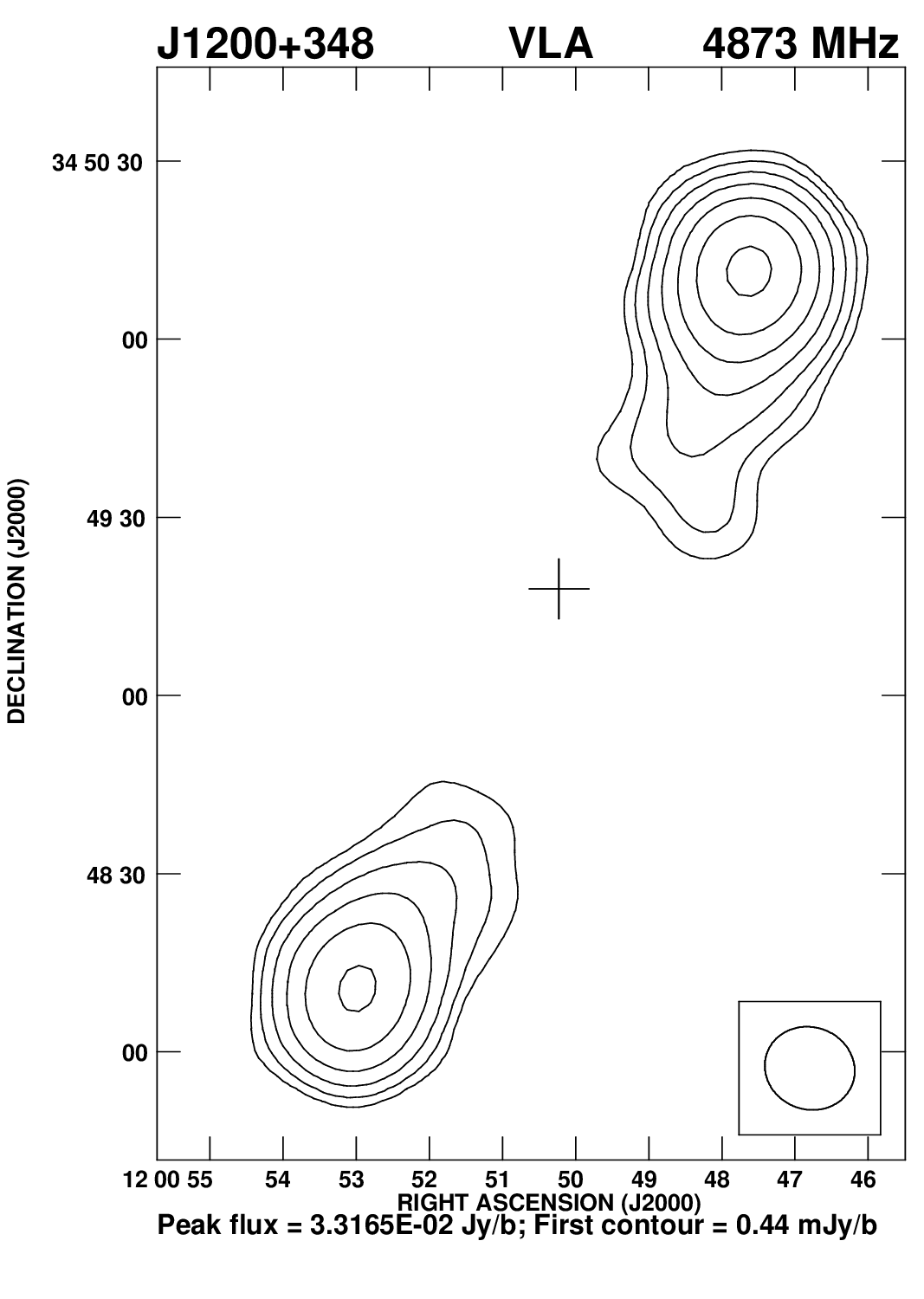,width=1.8in,angle=0}
  }
}
\contcaption{}
\end{figure*}

\subsection{Notes on the sources}
\noindent
{\bf J0135+379, 3C46} VLA B- and C-array images at L-band
(Gregorini et al. 1988; Vigotti et al. 1989) show the extended
lobes of emission. Higher-resolution A-array observations by
Neff et al. (1995) list the core flux density as 11 and 3.6 mJy
at 1478 and 4848 MHz, suggesting it might be a steep-spectrum core.
The $\lambda$6cm value is similar to the value of 2.3 mJy estimated by
Giovannini et al. (1988), but no significant core emission is seen in
the tapered image of Neff et al. which has an rms noise of 0.4 mJy/beam.  The
$\lambda$20cm value is not consistent with the estimate of 3 mJy by
Gregorini et al. (1988). We made an image using the longer spacings, and
find the core flux density to be $\sim$1.2 mJy at 4841 MHz and $\lapp$1.1 mJy
at 1465 MHz, yielding a spectral index $\gapp$0.1. In Fig. 1, we present the
$\lambda$6cm image and the spectral
index image between L and C bands superimposed on it. The spectral index,
$\alpha$ varies from $-$0.9 to $-$1.6 in the western
lobe, and from $-$1.0 to $-$1.8 in the eastern one. The implied ages
using the formalism of Myers \& Spangler (1985) are $\sim$2.3$\times$10$^7$
and $\sim$3.0$\times$10$^7$ yr respectively.

\noindent
{\bf J0139+399, 4C39.04} The large-scale structure showing the
relaxed lobes has been reported earlier
by a number of authors (e.g. Hine 1979; Vigotti et al. 1989). The flux densities
listed in Table 2 suggest that the spectral index of the entire source between
1.3 and 4.8 GHz is $-$1.34 while that of the core is $-$0.8. The existence of an
SSC was noted earlier (e.g. Hine 1979; Klein et al. 1995).
Similar-resolution observations of about 5 arcsec between 1.4 and 5 GHz
(Fomalont \& Bridle 1978; Gregorini et al. 1988; Bondi et al. 1993;
Klein et al. 1995) also yield a core spectral index of $\sim-$0.8.  This
is consistent with the 10.6 GHz value of
4$\pm$1 mJy (Mack et al.  1994). Saripalli et al. (1997) find the core
to be a compact source with a flux density of 20$\pm$5 mJy from VLBI
observations at 1.67 GHz with a resolution of 25 mas. Our GMRT image
highlights the diffuse lobes and bridge of emission, and
well resolves the possibly unrelated source to the north-east 
(RA 01$^h$ 39$^m$ 46.$^s$9, Dec 39$^\circ$ 58$^\prime$ 11.$^{\prime\prime}$3), 
which appears to be an FRI type of source with a flux density of 42 mJy at 1287 MHz.
The spectral index,
$\alpha$ varies from $-$1.1 to $-$1.8 in the western
lobe, and from $-$0.7 to $-$1.5 in the eastern one, suggesting that the ages
are $\sim$3$\times$10$^7$ and $\sim$4.5$\times$10$^7$ yr respectively.
Our values of core flux density are
similar to those measured by Gregorini et al. (1988) and Bondi et al. (1993) at the
L and C bands respectively, suggesting that the core has not varied over a time scale of $\gapp$10 yr.

\begin{table*}
\caption{ The observational parameters and observed properties of the sources}

\begin{tabular}{l l rrr r r r rr l rr r rr}
\hline
Source   & Freq.       & \multicolumn{3}{c}{Beam size}                    & rms      & S$_I$   & Cp  & S$_p$  & S$_t$  & Cp   & S$_p$ & S$_t$ & Cp  & S$_p$   & S$_t$     \\

           & MHz         & $^{\prime\prime}$ & $^{\prime\prime}$ & $^\circ$ &    mJy   & mJy     &     & mJy    & mJy    &      & mJy   & mJy   &     & mJy     & mJy       \\
           &             &                   &                   &          &  /b      &         &     & /b     &        &      & /b    &       &     & /b      &           \\ 
\hline
J0135+379  &  V1465      & 14.0    & 11.5       &  34                       &   0.18   & 1217    & W   & 218    & 640    & C    &$\lapp$1.1 &   & E   &  199    & 579       \\
           &  V4841      & 13.5    & 11.2       & 165                       &   0.12   &  356    & W   &  74    & 184    & C$^g$&      1.2  &1.1& E   &  59     & 172       \\

J0139+399  & G1287       &  6.2    &  3.8       &  37                       &   0.05   & 1201    & W   & 4.5    & 600    & C$^g$&  34   & 35    & E   &  10     & 579       \\
           & V4841       & 18.0    & 13.8       & 101                       &   0.16   &  203    & W   & 7.6    &  85    & C$^g$&  11   & 12    & E   & 9.4     & 110       \\

J0200+408  & V4841       & 13.0    & 11.5       & 146                       &   0.13   &         & W   &        &        & C$^g$& 2.8   & 2.6   & E   & 5.9     & 5.8       \\

J0313+413  &  G606       & 11.8    &  6.6       & 137                       &   0.33   &         & Jet &  20    & 179    & C$^g$& 348   & 375   & N   & 8.8     & 129       \\
           &  V1425      & 19.5    & 15.9       &  39                       &   0.10   &         & Jet &  13    &  63    & C$^g$& 302   & 308   & N   & 5.5     &  21       \\

J0657+481  &  G617       & 28.1    & 10.3       &  90                       &   0.81   &  259    & W   &  74    & 152    &      &       &       & E   &  53     &  94       \\
           &  V1425      & 13.6    &  8.0       &  75                       &   0.10   &   73    & W   &  23    &  40    & C    &$\lapp$0.6 &   & E   &  19     &  29       \\
           &  V4848      & 15.9    & 12.8       & 131                       &   0.10   &   22    & W   & 8.1    &  10    & C$^g$& 0.7   & 0.9   & E   & 7.0     &  9.3      \\

J0754+432  & V4816       & 16.8    & 12.8       &  82                       &   0.10   &   38    & N   & 1.9    &  11    & C$^g$&  16   &  16   & S   & 1.0     & 7.4       \\

J0819+756  & V4841       & 20.5    & 11.1       &  60                       &   0.10   &  181    & W   &  19    & 106    & C$^g$&  46   &  46   & E   & 4.6     & 30        \\

J1101+365  &  V1425      & 15.2    &  5.0       & 104                       &   0.10   &  111    & N   &  17    &  48    & C?   &$\sim$0.6 & $\sim$0.8  & S   &   23    &  60       \\
           &  V4841      & 24.1    & 14.7       & 107                       &   0.10   &   38    & N   &  11    &  17    & C?   &$\sim$0.6   &$\sim$0.6   & S   &  12     &  20       \\

J1200+348  &  V1465      & 15.7    &  5.0       & 109                       &   0.10   &  198    & N   &  65    & 132    & C    &$\lapp$0.2 &   & S   &   20    &  64       \\
           &  V4873      & 15.3    & 13.8       &  70                       &   0.11   &   77    & N   &  33    &  50    & C    &$\lapp$0.3 &   & S   &  16     &  26       \\

J1235+213  &  V1385      & 16.9    &  12.1      & 103                       &   0.27   & 2975    & W   & 543    & 1549   &      &           &   & E   &   586   &  1431     \\

J1313+696  &  G605       &  8.1    &  4.6       & 151                       &   0.20   & 2499    & N   &  39    & 1441   & C    &$\lapp$2.0 &   & S   &  27     &  1046     \\
           &  V1425      & 14.9    &  6.5       & 121                       &   0.10   & 1383    & N   &  54    & 797    & C$^g$& 4.6   & 7.0   & S   &  40     &  585      \\
           &  V4873      & 20.4    & 12.2       &  60                       &   0.13   &  431    & N   &  30    & 255    & C$^g$& 4.1   & 4.3   & S   &  18     &  172      \\

J1604+375  &  G613       &  7.8    &  4.8       & 171                       &   0.10   &  264    & N   &  14    & 111    & C$^g$& 5.4   & 6.4   & S   &  53     &   142     \\
           &  V1425      & 32.3    & 10.4       &  58                       &   0.21   &  106    & N   &  17    &  43    & C$^g$& 1.9   & 2.9   & S   &  31     &   56      \\
           &  V4873      & 37.9    & 11.8       & 116                       &   0.14   &   25    & N   & 4.9    & 7.5    & C$^g$& 1.4   & 1.5   & S   & 9.3     &   14      \\

J1637+417  &  V1425      & 23.0    &  8.6       &  54                       &   0.10   &   52    & N   & 5.1    &  14    & C$^g$& 3.9   & 3.9   & S   &   19    &  34       \\
           &  V4873      & 41.3    & 11.8       & 119                       &   0.12   &   19    & N   & 2.0    & 4.7    & C$^g$& 2.3   & 2.1   & S   & 9.6     &  11       \\

J1702+422  & G602        & 15.6    & 14.0       &  26                       &   0.26   &  393    & N   &  56    & 235    & C    &$\lapp$1.7 &   & S   &  39     &   159     \\
           & V1425       & 21.7    &  6.7       &  53                       &   0.11   &  172    & N   &  26    & 102    & C    &$\lapp$0.6 &   & S   &  17     &   72      \\
           & V4848       & 45.1    & 12.0       & 120                       &   0.11   &   51    & N   &  12    &  31    & C$^g$& 1.7   & 1.7   & S   & 6.5     &   18      \\

J1919+517  & V4866       & 24.7    & 18.7       &  74                       &   0.23   &   70    & N   &  23    &  53    & C$^g$& 5.3   & 5.7   & S   & 4.2     &   14      \\

J2042+751  & G599        & 29.5    & 14.2       & 143                       &   1.80   & 4078    & N   &  82    &1208    & C$^g$& 152   & 174   & S   & 1001    &  2623     \\
           & V4816       & 20.0    & 10.9       & 110                       &   0.16   &  471    & N   & 5.1    &  66    & C$^g$& 250   & 249   & S   &  76     &   172     \\

J2312+187  & V1425       & 14.2    & 12.7       & 179                       &   0.26   & 1885    & N   & 311    & 802    & C$^g$& 3.2   & 4.1   & S   &  356    &  1069     \\
           & V4866       & 14.2    & 13.7       & 120                       &   0.12   &  538    & N   & 102    & 230    & C$^g$& 2.9   & 3.6   & S   & 123     &  307      \\
\hline

\end{tabular}
\end{table*}

\noindent
{\bf J0200+408, 4C40.08} It has been classified as a double source with
no detected radio core (Vigotti et al. 1989; Gregorini et al. 1988;
Schoenmakers et al. 2000a). The NVSS image (Fig. 1) shows the
diffuse large-scale structure along a PA of 110$^\circ$ 
with no prominent hotspot at the outer edges.  Our VLA observations at 4.8 GHz reveal 
a radio core which is coincident with the optical identification and has 
a flux density of 2.6 mJy. We also detect the compact
component to the east which has a spectral index of $\sim-$0.6 using the
NVSS flux density at 1.4 and our value at 5 GHz. An examination of the Digital Sky
Survey (DSS) image shows a faint galaxy coincident with the radio position of
this compact component, suggesting that this is an unrelated source.

\begin{figure}
\vbox{
    \psfig{file=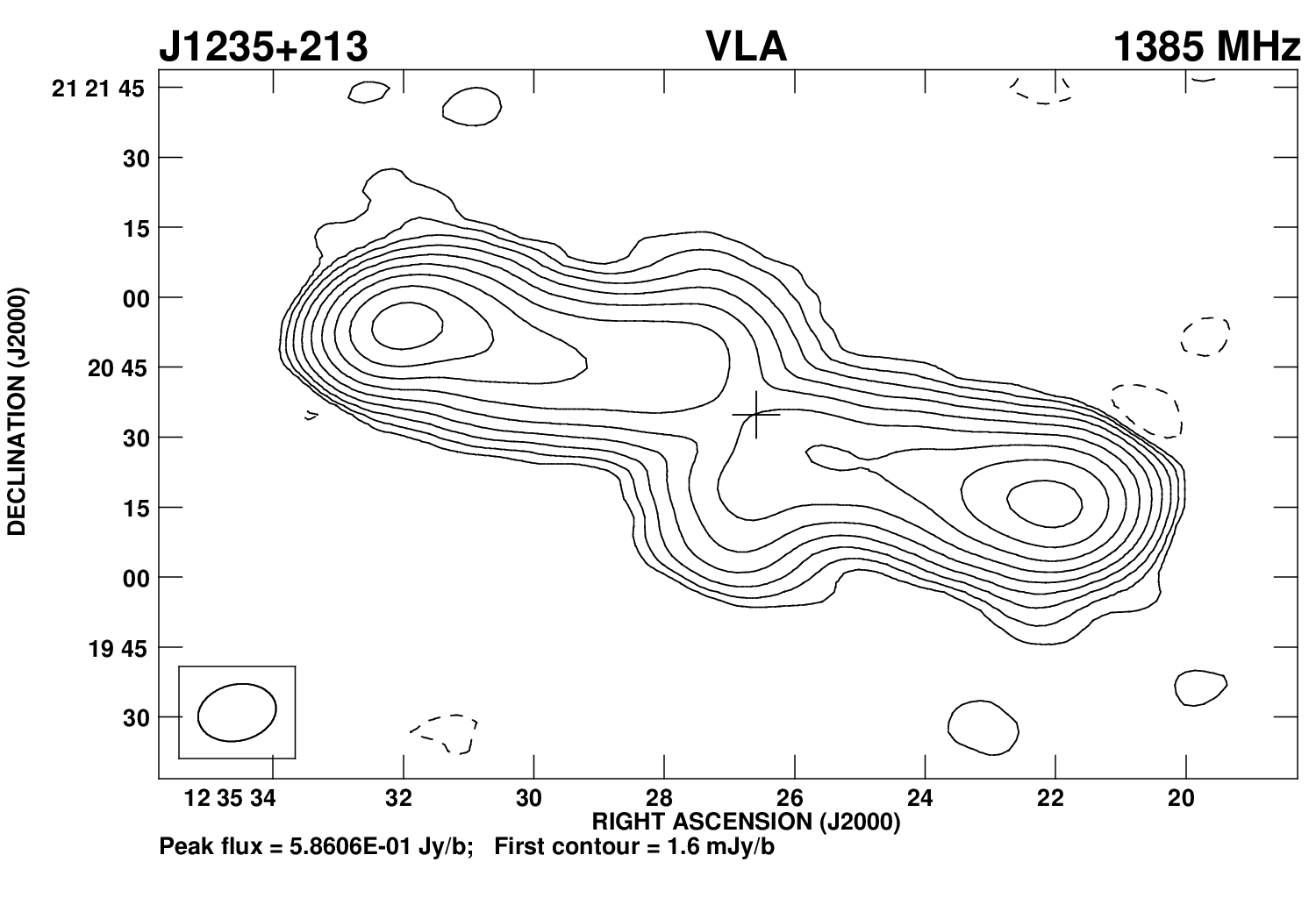,width=3.4in,angle=-90}
    }
\contcaption{}
\end{figure}

\begin{figure}
\vbox{
    \psfig{file=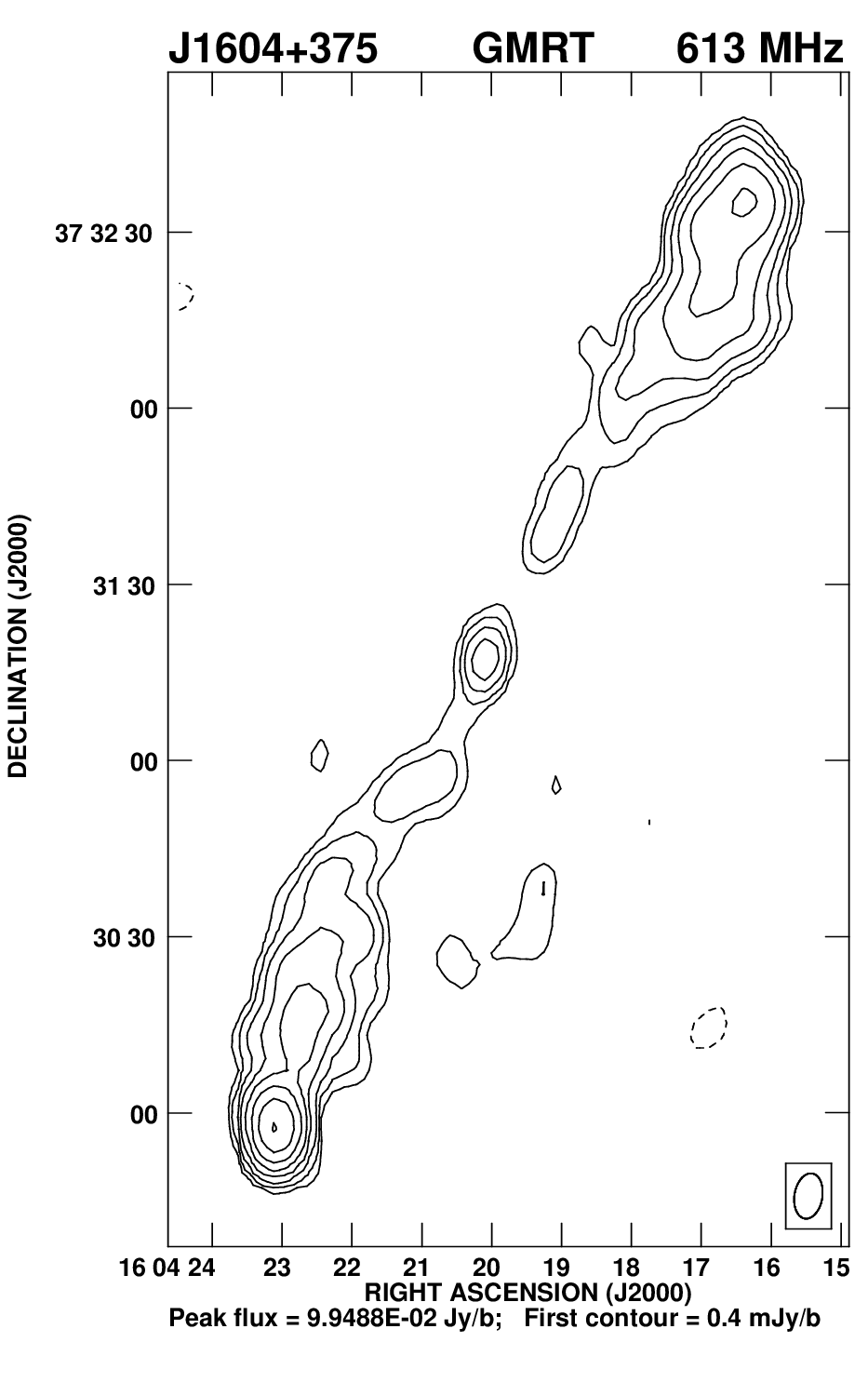,width=3.0in,angle=0}
    }
\contcaption{}
\end{figure}

\begin{figure*}
\vbox{
\hbox{
    \psfig{file=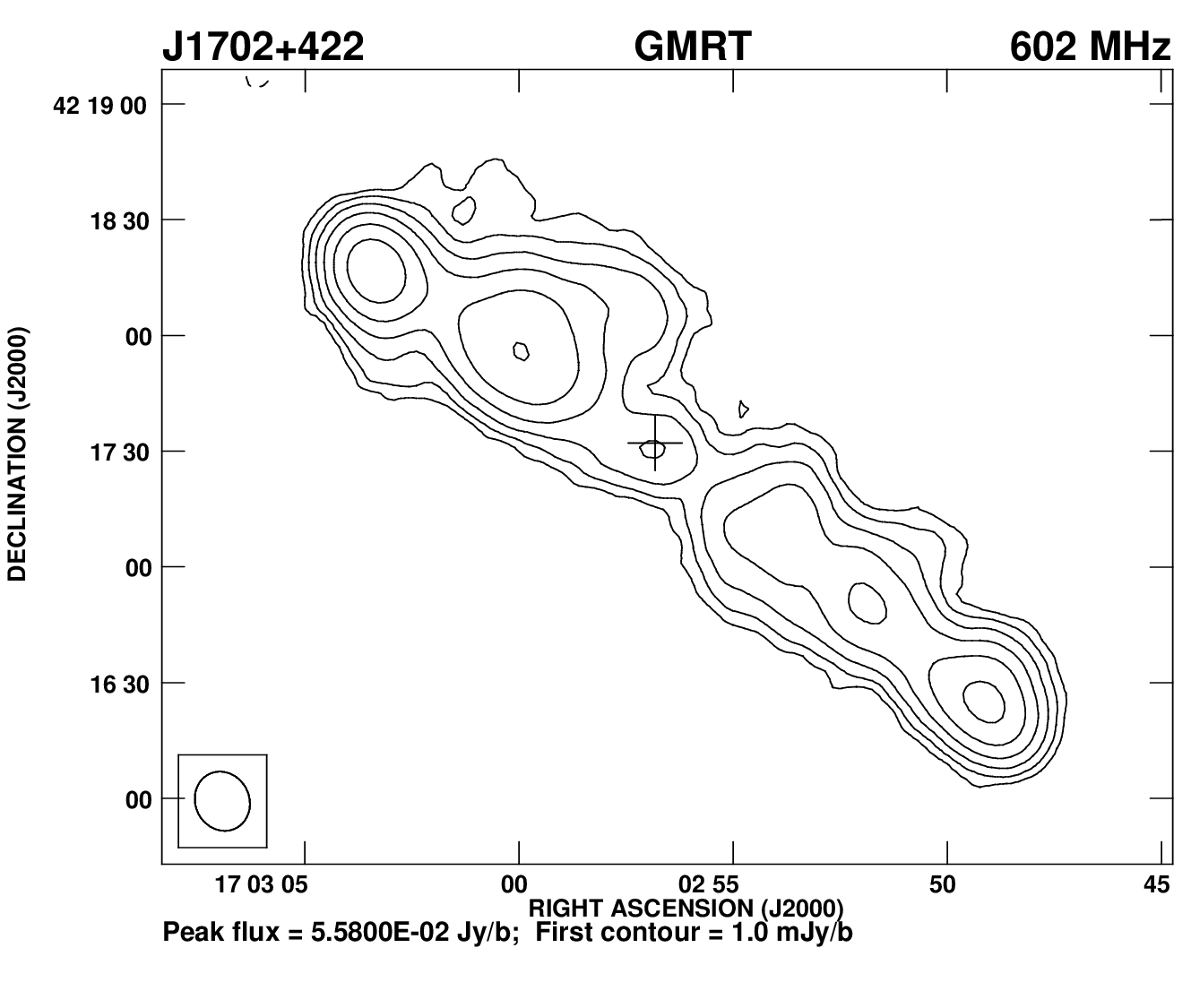,width=3.5in,angle=-90}
    \psfig{file=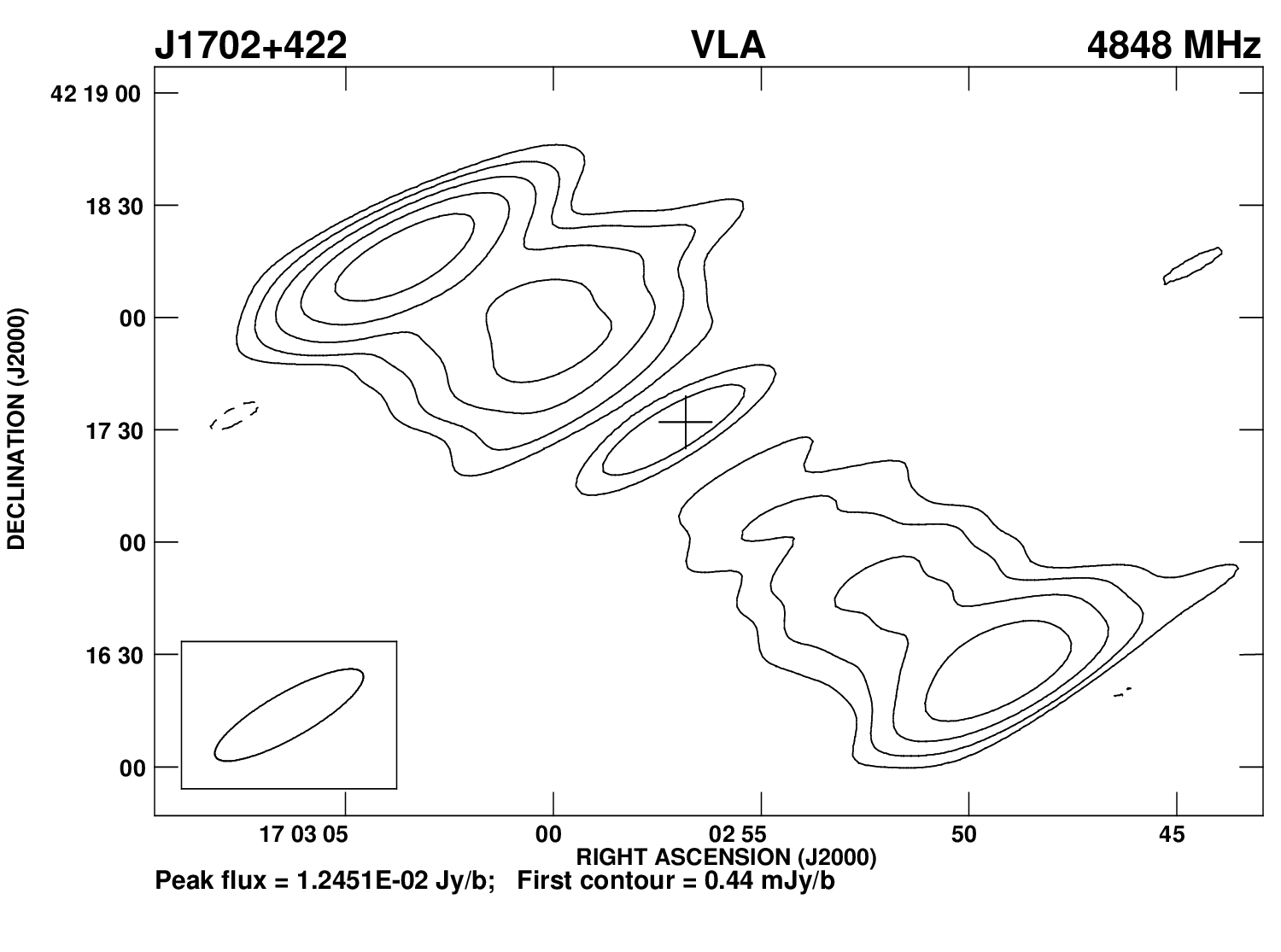,width=3.5in,angle=-90}
    }
\hbox{
    \psfig{file=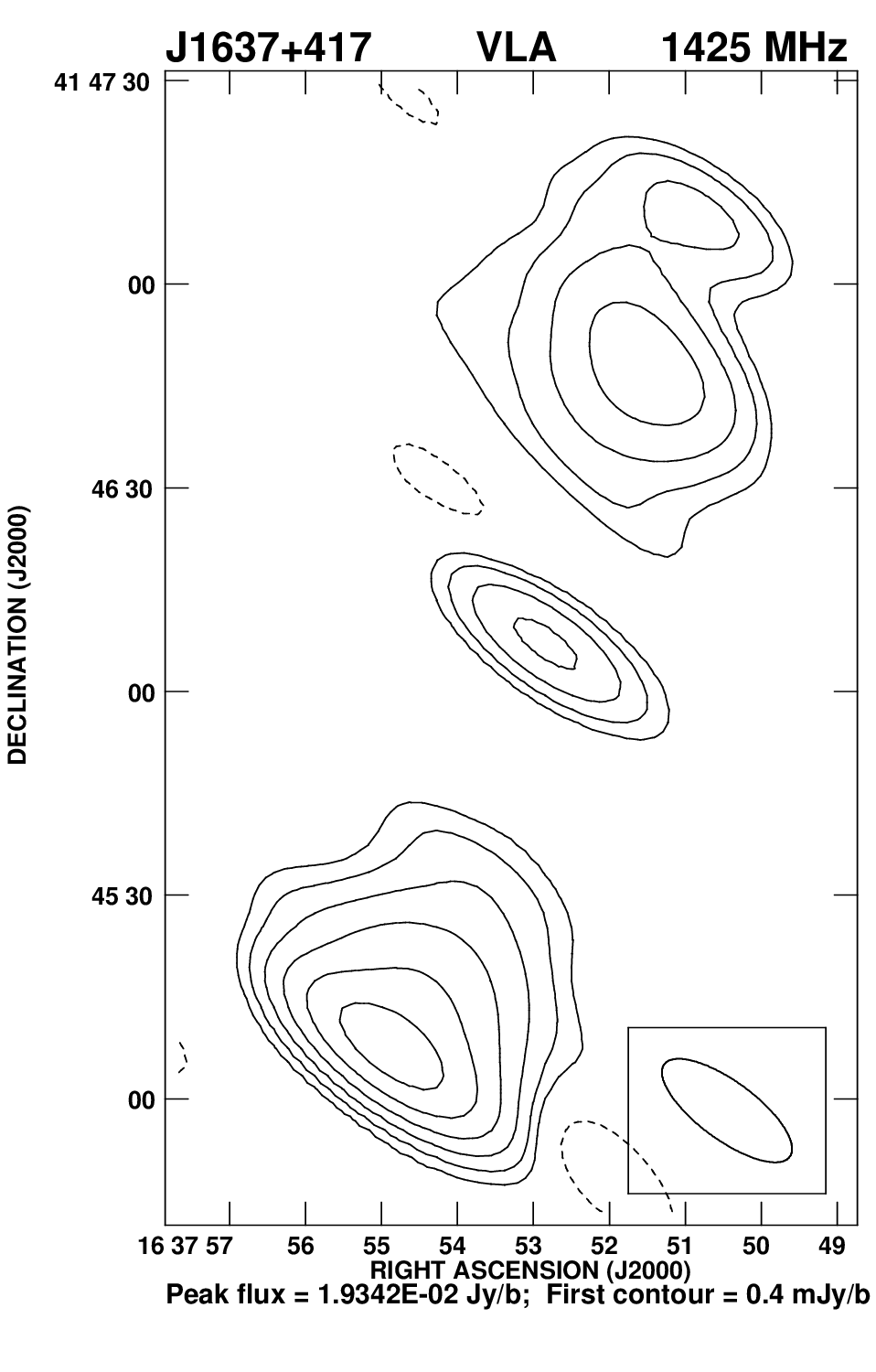,width=2.00in,angle=0}
    \psfig{file=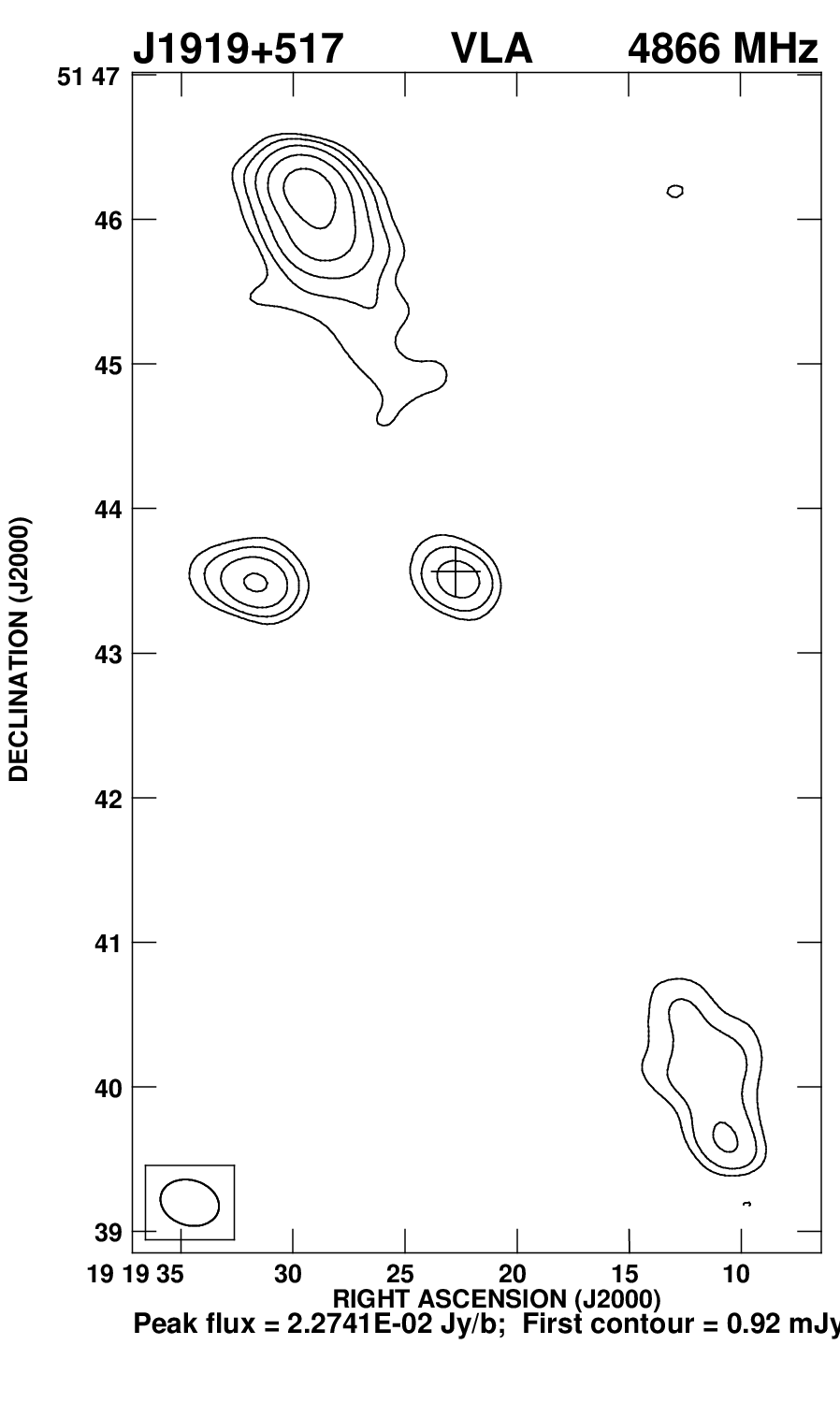,width=2.00in,angle=0}
    \psfig{file=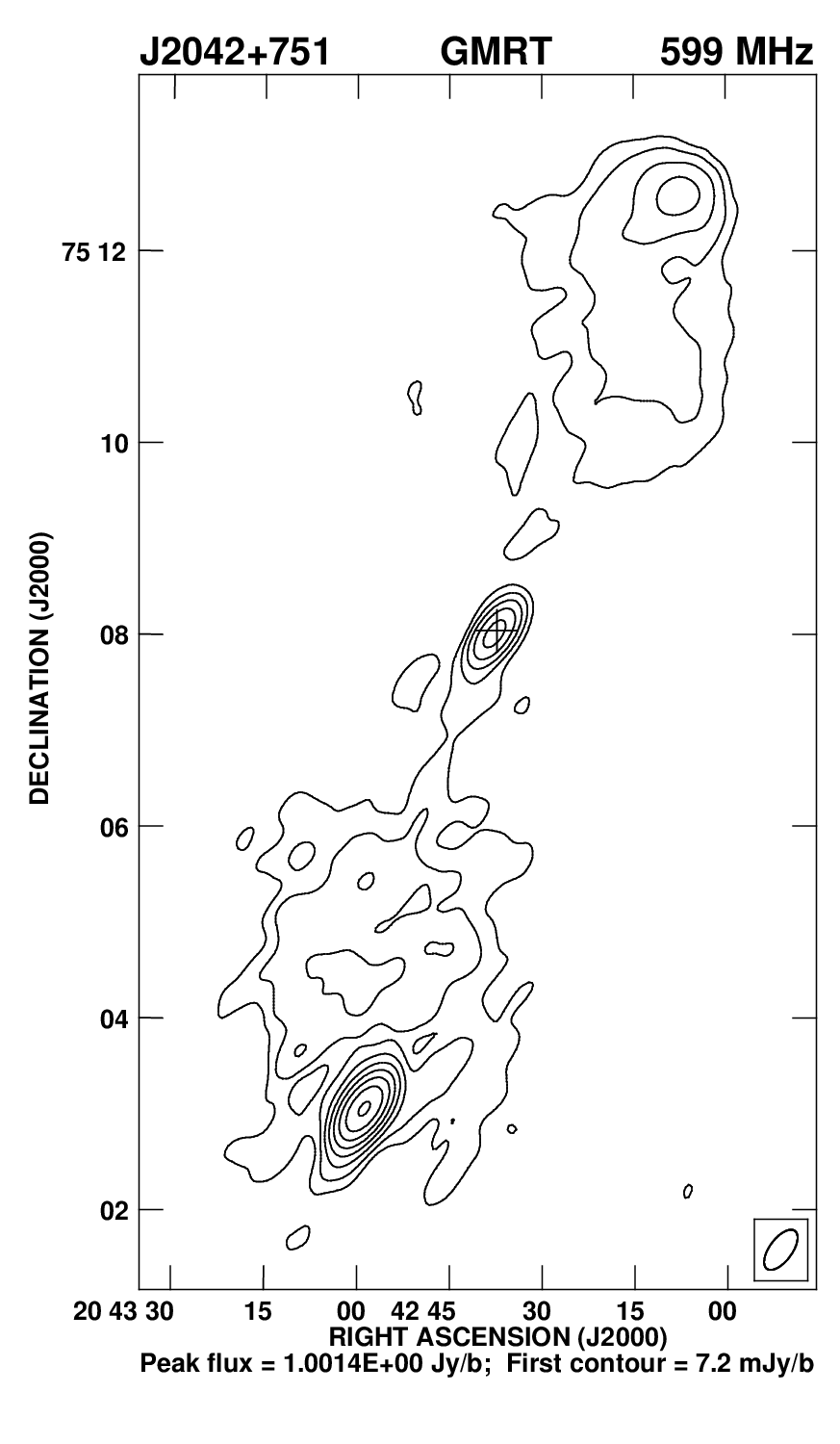,width=2.00in,angle=0}
     }
}
\contcaption{}
\end{figure*}

\noindent
{\bf J0313+413} The large-scale structure of this broad line radio galaxy
(March\~{a} et al. 1996) was presented by de Bruyn (1989) with rather
coarse resolution. Subsequent published observations with the VLA
and VLBI have revealed a prominent core and a weak jet-like extension
(Patnaik et al. 1992; Henstock et al. 1995; Taylor et al. 1996;
Schoenmakers et al. 2000a; Fey \& Charlot 2000).  Our GMRT image at
606 MHz reveals a curved radio jet extending over a distance of 250 kpc,
and then bending abruptly to enter the northern hotspot. Convolving
the GMRT image to that of our VLA image at 1425 MHz shows the core to
have a flat radio spectrum of $-$0.21 while in the jet the average value
of $\alpha$ is $\sim-$1.1, with the values close to the peaks of emission
being $\sim-$0.9.

\noindent
{\bf J0657+481, 7C} The GMRT image at 617 MHz shows the bridge of emission 
(Konar et al. 2003) while the VLA 4.8-GHz image shows the radio core co-incident with the
galaxy. The core has an inverted spectrum with a spectral index $\gapp$0.3 between 1.4 and
4.8 GHz.  The component to the north-west is possibly unrelated. This GRS was identified 
by Cotter et al. (1996) from the 7C survey.

\noindent
{\bf J0754+432} The WENSS, NVSS and FIRST images have been presented by
S99. Our 4.8-GHz VLA image shows the core to have
a flat spectrum ($\alpha\sim$0.07), and the southern lobe to consist of 
three features which are also visible in the FIRST image. The spectral indices
of the western, southern and eastern features using the flux densities from the
FIRST and our images are
$-$0.51, $-$1.0 and $-$0.58 respectively. The weaker features are more compact
and significantly flatter, suggesting that these are possibly unrelated to the GRS. 

\noindent
{\bf J0819+756} The WENSS and NVSS images have been 
presented by S99, while a VLA L-band image with a resolution
of 8 arcsec
has been published by Lara et al. (2001). Our core flux density at 4.8 GHz
is 46 mJy,  which is much lower than the value of 57 mJy found by Lara et al. at 4.9 GHz,
suggesting variability of the core flux density. This would be consistent with its
prominent core and a small angle of inclination to the line of sight. This giant source has
multiple hotspots; the furthest component towards the north-east being an
unrelated source (cf. Lara et al. 2001).

\noindent
{\bf J1101+365, 7C}
This GRS has been identified from the 7C survey by Cotter et al. (1996).
We detect a possible core which has a flat radio spectrum between 1.4 and 4.8 GHz.

\noindent
{\bf J1200+348}    
A VLA C-array image at 1465 MHz of this GRS from the GB/GB2 sample has been
published by Machalski \& Condon (1985).

\noindent
{\bf J1235+213, 3C274.1} The core flux density has an average value of
$\sim$12 mJy at 963 MHz (Bedford et al. 1981; Kerr et al. 1981) and 
3 mJy at 4874 MHz (Strom et al. 1990), yielding a core spectral index of $-$0.85.

\noindent
{\bf J1313+696, 4C69.15} The 1.4-GHz image shows the prominent bridge of
emission, with a weak radio core. The limit on the core flux density at 605 MHz
suggests that the core spectrum turns over at $\lapp$1 GHz. The core spectral
index between 1.4 and 4.8 GHz is $-$0.4.

\noindent
{\bf J1604+375, 7C} The GMRT 613-MHz image shows possible evidence of a
curved twin-jet structure, with the overall structure resembling an S-shaped source.
Such structures could arise due to precession of the jet axis.

\noindent
{\bf J1637+417, 7C} The core spectral index from our VLA observations is $\sim-$0.50,
suggesting that it could be an SSC.

\noindent
{\bf J1702+422, 7C} The GMRT 602-MHz image shows a prominent bridge of emission
with no core component. Our estimate of the core flux density  of $\lapp$0.6 mJy at 
1.4 GHz is consistent with a weak feature of 0.66 mJy seen in the FIRST image. 
Our 4.8-GHz image shows a core with a flux density of 1.7 mJy showing that it has 
an inverted radio spectrum ($\alpha\gapp$0.8). 

\noindent
{\bf J1919+517} The WENSS and a Westerbork L-band image have been presented
by S99, and it has been observed with low resolution
at 10.7 GHz with the Effelsberg telescope (Mack et al. 1997 and references therein).
Our VLA 4.8-GHz image clearly shows the radio core with a flux density of 5.7 mJy,
which combined with the 1.4 GHz value listed by S99 yields
a spectral index of $\sim-$1.0. The component east of the core is possibly unrelated
and has a flux density of 9.2 mJy at 4866 MHz.

\begin{figure}
\vbox{
    \psfig{file=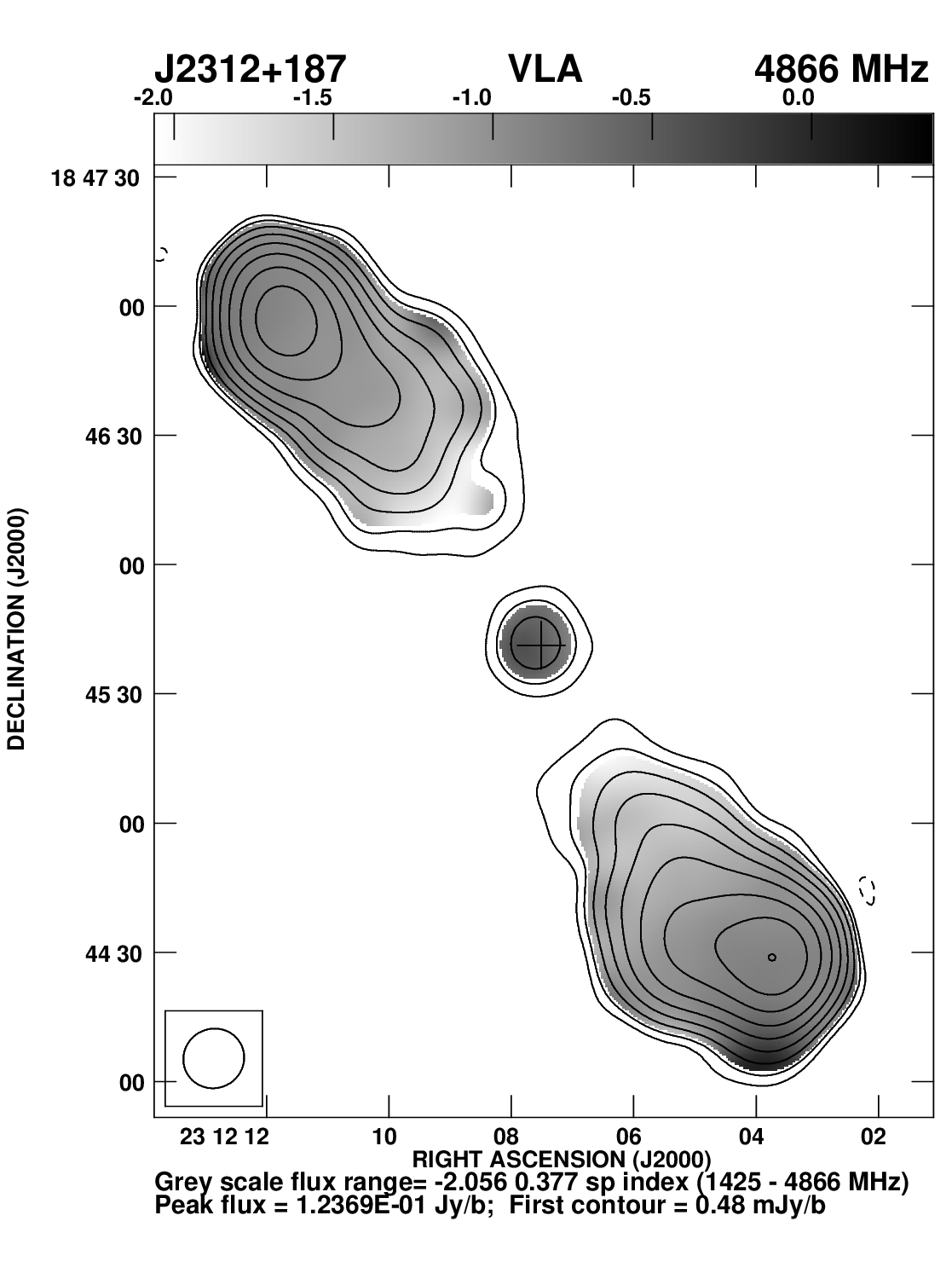,width=3.00in,angle=0}
    }
\contcaption{}
\end{figure}

\noindent
{\bf J2042+751, 4C74.26} This GRS, identified with a quasar has a prominent core and
a one-sided radio jet (Riley et al. 1989; Riley \& Warner 1990; Pearson et al. 1992). 
The total flux density of the core at 5 GHz decreased from 0.42 to 0.31 Jy between 
1986 and 1988 (Riley et al. 1989). Our estimate of 250 mJy at 4816 MHz in 2000 is 
consistent with a strongly variable radio core.

\noindent
{\bf J2312+187, 3C457} The spectral indices of both the northern and sourthern lobes
between 1.4 and 4.9 GHz varies from $\sim-$0.9 to $-$2.0, suggesting a spectral age
of $\sim$3.5$\times$10$^7$ yr. The region with a flatter spectral index at the edge
of the southern lobe is close to the position of the unrelated source identified by
Leahy \& Perley (1991). The core spectral index is $-$0.1 between 1.4 and 4.9 GHz.

\begin{table*}
\caption {Physical properties of the sources} 

\begin{tabular}{l r c l c r c r c r r c r r}
\hline
Source &  Alt.  & Opt.& Redshift&LAS&{\it l}&P$_{1.4}$ & f$_{\rm c}$ & Cmp.  &  u$_{min}$ & B$_{eq}$& Cmp.  &u$_{min}$ & B$_{eq}$ \\
       & name   & Id. &         &      &       &          &      & & 10$^{-14}$ &         &       & 10$^{-14}$ &       \\
       &        &     &  &$^{\prime\prime}$&kpc&W Hz$^{-1}$ &    & &  J m$^{-3}$&   nT    &       & J m$^{-3}$ &   nT      \\
\hline
J0135+379&   3C46  &  G  & 0.4373  &  150 &  846  & 27.01    &$\sim$0.004 &   W    &    26.36   &  0.53   & E & 20.67 &0.47    \\
J0139+399& 4C39.04 &  G  & 0.2107  &  370 & 1259  & 26.15    &      0.07  &   W    &    19.30   &  0.46   & E & 11.83 &0.36    \\
J0200+408& 4C40.08 &  G  & 0.0827  &  920 & 1414  & 24.61    &      0.05  &   W    &     1.09   &  0.11   & E &  1.40 &0.12    \\
J0313+413&    B3   &  G  & 0.136   &  570 & 1357  & 25.34    &      0.86  &   Jet  &    44.44   &  0.69   &   &       &        \\
J0657+481&    7C   &  G  & 0.776   &  133 &  989  & 26.34    &      0.04  &   W    &    14.57   &  0.40   & E & 14.55 &0.40    \\
J0754+432&         &  G  & 0.3474  &  485 & 2368  & 25.67    &      0.35  &   N    &     0.99   &  0.10   & S &  0.55 &0.08    \\
J0819+756&         &  G  & 0.2324  &  454 & 1666  & 25.97    &      0.31  &   W    &     3.39   &  0.19   & E &  2.69 &0.17    \\
J1101+365&    7C   &  G  & 0.750   &  154 & 1131  & 26.49    &$\sim$0.01  &   N    &     4.25   &  0.21   & S &  4.35 &0.22    \\
J1200+348&         &  G  & 0.55    &  145 &  927  & 26.43    &$\lapp$0.006&   N    &     8.01   &  0.29   & S &  5.75 &0.25   \\
J1235+213& 3C274.1 &  G  & 0.422   &  151 &  834  & 27.28    &      0.004 &   W    &    24.45   &  0.51   & E & 26.48 &0.53   \\
J1313+696& DA340   &  G  & 0.106   &  388 &  745  & 25.60    &      0.01  &   N    &     6.54   &  0.27   & S &  5.24 &0.24   \\
J1604+375&    7C   &  G  & 0.814   &  178 & 1346  & 26.54    &      0.04  &   N    &    45.55   &  0.70   & S & 21.42 &0.48   \\
J1637+417&    7C   &  G  & 0.867   &  134 & 1035  & 26.36    &      0.09  &   N    &     3.65   &  0.20   & S &  5.47 &0.24  \\
J1702+422&    7C   &  G  & 0.476   &  196 & 1160  & 26.16    &      0.03  &   N    &     5.04   &  0.23   & S &  6.88 &0.27  \\
J1919+517&         &  G  & 0.284   &  429 & 1825  & 25.95    &      0.06  &   N    &     4.40   &  0.22   & S &  3.54 &0.20  \\ 
J2042+751& 4C74.26 &  Q  & 0.104   &  610 & 1151  & 25.68    &      0.42  &   N    &     2.73   &  0.17   & S &  3.83 &0.20  \\ 
J2312+187&  3C457  &  G  & 0.427   &  190 & 1056  & 27.09    &      0.006 &   N    &    18.34   &  0.44   & S & 19.30 &0.46  \\    
\hline

\end{tabular}
\end{table*}

\section{Discussion and results}
\subsection{Radiative losses}
For the sources listed in Table 2 we estimate the minimum energy density, u$_{min}$
and the equipartition magnetic field, B$_{eq}$ for all the extended components
(Miley 1980; Longair 1994). These are listed in Table 3 which is arranged as follows.
Columns 1 and 2: source name and an alternative name.  Column 3:
optical identification where G denotes a galaxy and Q a quasar;
column 4: redshift; columns 5 and 6: the largest angular size in arcsec and 
the corresponding projected linear size in kpc; 
column 7: the luminosity at 1.4 GHz in logarithmic units of W Hz$^{-1}$;
column 8: the fraction of emission from the core, f$_c$,  at an emitted 
frequency of 8 GHz. For sources without detailed spectral information,
spectral indices of 0 and $-$0.9 have been assumed for the cores and lobes
respectively. Columns 9 and 12: component designation; columns 10, 11 and 13, 14: the 
corresponding minimum energy density, u$_{min}$, in units of 10$^{-14}$ J m$^{-3}$ and the 
equipartition magnetic field, B$_{eq}$, in nT (1T = 10$^{4}$ G) for the extended components.
For sources such as J0200+408, where most of the lobe emission has been resolved out
in our images, we have used the NVSS images for estimating these parameters. 
The minimum energy density and equipartition magnetic field have been estimated
for the extended emission assuming a cylindrical or
spheroidal geometry, a filling factor of unity and that
energy is distributed equally between relativistic electrons and protons. 
The size of the lobes has been estimated from the lowest contours in the 
available images.  The luminosity has been estimated between 10 MHz
and 100 GHz using the known spectral indices of the extended emission or
else assuming a value of $-$0.9. 

The minimum energy densities are in the range
of $\sim$1 to 45 $\times$ 10$^{-14}$ J m$^{-3}$ with a median value of about
5.8 $\times$ 10$^{-14}$ J m$^{-3}$, while the equipartition magnetic
field for the lobes range from 0.08 to 0.7 nT with a median value of 0.25 nT. 
We examine the relative importance of synchrotron and
inverse-Compton losses in the lobes of these radio sources.
The equipartition magnetic field of the lobes for all the 
sources are less than the equivalent magnetic field of the microwave
background radiation at the redshift of the source, B$_{ic}$=0.32(1 + z)$^2$ nT.
This suggests that the inverse-Compton losses are larger than
the synchrotron radiative losses in the evolution of the lobes of these giant
sources, consistent with earlier results (e.g. IC99). 

The dominance of inverse Compton losses
is likely to severely affect the appearance and identification of GRSs at 
high redshifts due to the suppression of bridge emission by inverse-Compton 
losses against the cosmic microwave background radiation, which increases sharply
with redshift. This could lead to `tail-less' hotspots leading to
their classification as independent radio sources (e.g. Baldwin 1982). 
For the sources discussed here, we quantify the prominence
of the bridge emission, f$_{\rm bridge}$, as the ratio of emission from the
bridge to that of the total emission, estimated from either single-dish
or low-resolution observations available in the literature. The flux density from the
bridge has been estimated by subtracting the `hot-spot' and core flux densities 
from the total flux density. The `hot-spot' flux densities have been estimated
by smoothing our images to a uniform linear resolution of $\sim$70 kpc and taking
the peak flux densities at the outer edges as a measure of the `hot-spot' flux densities.
It was possible to do this for 14 of the objects in our sample. 
With observations of higher linear resolution, one could get better estimates of
the `hot-spot' flux densities. A plot of f$_{\rm bridge}$ at an emitted frequency
of 1.4 GHz against redshift is shown in Fig. 2. The error bars indicate a $\pm$1$\sigma$
error in f$_{\rm bridge}$ adopting an error of 10 per cent in the flux densities. 
The plot clearly shows an inverse correlation, with a Spearman rank correlation 
coefficient of 0.52, corresponding to a confidence level of $>$95 per cent, suggesting 
that suppression of bridge emission could affect the identification of GRSs at high redshifts.

\subsection{Relic radio emission}
In order to understand the nature of episodic activity in the nucleus of the
parent galaxy, which sometimes manifests itself as a double-double radio galaxy 
(Schoenmakers et al. 2000b), we attempt to identify relic radio emission from an earlier
cycle of nuclear activity.
The lifetime of a relativistic electron at an observed frequency, $\nu_o$, 
due to both synchrotron and inverse-Compton losses is given by 
$${\rm \tau =
{5.03\times 10^4 \over [(1+z)\nu_o]^{1/2}B_{eq}^{3/2}[1 + (B_{ic}/B_{eq})^2]} \ \ Myr}, $$
where $\nu_o$ is in MHz and the magnetic fields B$_{ic}$ and B$_{eq}$ are in units of 10$^{-10}$ T
(e.g. Mack et al. 1998).
For a median redshift of $\sim$0.4 as in our sample of sources and a median magnetic field of 0.25 nT,
the lifetime of the relativistic electron at an observed frequency of 1400 MHz is 
$\sim$4$\times$10$^7$ yr. For a velocity of advancement of 0.1c, the outer hotspots would traverse
a distance of $\sim$1.2 Mpc during this time period.

Radio sources, with an inactive nucleus for timescales $\gapp\tau$, may be
characterised by a very relaxed lobe with no hotspot at the outer edges and a steep radio spectrum
($\alpha\lapp-$1.0). Detection of a radio core in such a source could indicate the beginning of a new 
period of activity. Such sources can be distinguished from the low-luminosity FRI sources.
Although FRI sources also have lobes without hotspots, these sources are characterised by radio jets 
which expand to form the diffuse lobes of emission, reflecting ongoing activity in the nucleus of the
parent galaxy. Also, the spectral indices of FRI sources are usually flatter, consistent with the 
radio luminosity$-$spectral index correlation (e.g. Laing \& Peacock 1980)

\begin{figure}
\vbox{
    \psfig{file=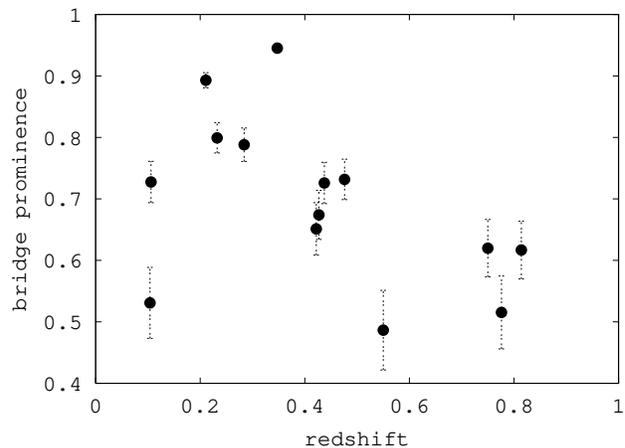,width=3.45in,angle=-90}
    }
\caption{The fraction of bridge emission, f$_{\rm bridge}$ at an emitted frequency of 1.4 GHz plotted against the redshift.}
\end{figure}

Two candidate sources with evidence of relic radio emission are J0139+399 and J0200+408.
J0139+399 has a very relaxed morphology with no hotspots, a weak radio core (Fig. 1), no jets and a spectral 
index of $-$1.34 from similar resolution GMRT and VLA data at L and C bands respectively. 
The second candidate, J0200+408, has similar structural features and 
a very steep spectrum of $-$1.78 using the flux density at 325 MHz
(S99) and the NVSS value at 1400 MHz. 
 
\subsection{Steep-spectrum cores (SSCs)}
The cores of extended double-lobed sources usually tend to have flat and complex radio spectra with
$\alpha\gapp-$0.5. The occurrence of SSCs with $\alpha\lapp-$0.5 
is indeed rare, and is often due to contamination of the measured low-frequency flux density by more
extended emission (e.g. Saikia, Kulkarni \& Porcas 1986). However, the occurrence of an SSC 
suggests that the emission is largely optically thin,  
which could be due to small-scale jets and/or lobes not resolved by the existing observations.
One of the well-known sources with an SSC is the GRS 3C236 where the core spectral index was 
estimated by Mack et al. (1997) to be $-$0.61$\pm$0.01. VLBI imaging of the SSC 
shows it to consist of a core, jet and two oppositely-directed lobes of emission 
(Schilizzi et al. 2001). The small- and large-scale structure of 3C236 suggests that it could be
considered to be a  double-double radio galaxy, which is a sign of recurrent activity.
It is interesting to note that 3C236 shows evidence of star formation
and also HI absorption against a lobe of the inner radio source (Conway \& Schilizzi 2000; Schilizzi et al. 
2001; O'Dea et al. 2001), which could be due to infall of gas fuelling the central engine.
Another possible steep spectrum core is associated with the giant radio galaxy DA240
and has a spectral index of $\sim-$0.56$\pm$0.01 (Mack et al. 1997). VLBI 
observations of the core with an angular resolution of 25 mas shows that it possibly has a 
twin-jet structure (Saripalli et al. 1997).

Of a sample of 30 radio galaxies, with known cores observed with the VLA A-array between 1.4 and 15 GHz, 
no radio core, except the GRS DA240, showed evidence of a straight, steep radio spectrum. The 
radio core of DA240 was found to have a flux density of 191 and 110 mJy at
1465 and 4885 MHz, yielding a spectral index of $\sim-$0.5 (Saikia et al., in preparation). Only three of
these 30 sources are GRSs, the other two being NGC315 and NGC6251, both of which have flat-spectrum cores. Similarly for a 
sample of 17 sources from the Molonglo Reference Catalogue observed with the VLA A-array between 1.4 and 15 GHz, 
the cores tend to have flat and complex spectra. Only two sources whose low-frequency flux densities were 
significantly affected by contamination from more extended emission appeared to have steep spectra (Ishwara-Chandra
\& Saikia 2000). Only two of these 17 sources are GRSs, the remaining ones being of smaller sizes.

Considering the rarity of SSCs and the occurrence of two of these with known GRSs, it is meaningful to enquire
whether this is entirely due to chance or it might have a deeper significance. For example, if the occurrence of SSCs
is related to recurrent activity, the preferential occurrence of SSCs in GRSs might enable us to get further insights
into the time scales of recurrent activity. Towards this end, we have tried to identify SSC candidates in the 
present sample of GRSs and have identified four candidates. As discussed in Section 3.1, J0139+399, J1235+213 and
J1919+517 appear to have $\alpha_{\rm core}\lapp-$0.8, while J1637+417 is a more marginal case with 
a steep spectrum, $\alpha_{\rm core}\sim-$0.5. For confirmation, 
the spectra of these candidate SSCs need to be determined over a larger 
frequency range from higher resolution observations. If confirmed, this would indicate a higher incidence of SSCs
among GRSs compared with smaller-sized sources as discussed earlier, which is possibly related to the time scales 
of recurrent activity.

\subsection{Core prominence and variability}
The fraction of emission from the core at an emitted frequency of 8 GHz, f$_{\rm c}$, ranges from 
$\lapp$0.004 to $\sim$0.86, with a median value of $\sim$0.04. The values of f$_{\rm c}$ are
usually comparable to the other sources of similar radio luminosity. The four sources with prominent
cores (f$_{\rm c}$$\gapp$0.2) are J0313+413, J0754+432, J0819+756 and J2042+751, the first three
of which are associated with galaxies and J2042+751 with a quasar. Given the prominence of the
cores, it is reasonable to assume that these galaxies lie close to $\sim$45$^\circ$ to the line of
sight, the dividing line between galaxies and quasars in the unified schemes (e.g. Barthel 1989).
The classification of J0313+413 as a broad line radio galaxy (March\~{a} et al. 1996), the detection
of broad hydrogen Balmer lines in J0754+432 and a broad component in the H$\alpha$ line of J0819+756
(Schoenmakers et al. 2001) are consistent with this interpretation. For an inclination angle of 
$\sim$45$^\circ$  their intrinsic sizes would increase to $\sim$1900, 3350 and 2350 kpc respectively.

The GMRT and VLA images of the radio galaxy J0313+413 shows a prominent radio jet which curves towards 
the north and then bends abruptly towards the west (Fig. 1). There is also a suggestion of a weak
counter-jet in the GMRT image at 606 MHz. The jet to counter-jet brightness ratio varies from
$\sim$7 to 20 along the length of the jet. If the asymmetry is due to relativistic beaming, this
imples a jet speed of $\sim$0.4$-$0.6c for an inclination angle of $\sim$45$^\circ$. We have also compiled
the core flux densities of J0313+413 in Table 4. 
Although the data are not homogeneous, measurements of a higher flux density with higher resolution
compared with coarser-resolution observations at a different epoch at both 5 and 8 GHz suggests that the core is
variable, which is consistent with the assumed angle to the line of sight. The giant
quasar J2042+751 (4C74.26) which has a prominent core, exhibits a one-sided radio jet and 
variability of the core flux density by approximately 50 per cent over a two-year time scale
(Riley \& Warner 1990). Our value of the core flux density at 4816 MHz is only 250 mJy compared with
previous measurements in the range of 310 to 420 mJy (Riley et al. 1989; Pearson et al. 1992), 
consistent with a strongly variable core.  
In the quasar 4C74.26, the inferred orientation angle of 
$\mbox{\raisebox{-0.1ex}{$\scriptscriptstyle \stackrel{<}{\sim}$\,}}$49$^\circ$ (Pearson
et al. 1992) implies an intrinsic size  
$\mbox{\raisebox{-0.1ex}{$\scriptscriptstyle \stackrel{>}{\sim}$\,}}$1525 kpc. 

\begin{table}
\caption{Core flux densities of J0313+413 }
\begin{tabular}{l r l r r r}

\hline
Teles-    &  resn.             &     Date  & Freq.   &  S    & Ref.   \\
cope      &  $^{\prime\prime}$ &           & MHz     &  mJy  &        \\
\hline

GMRT      &     8.8            & 2003Sep06      &     606 &   375 &  1    \\
VLA-A     &     2.0            & 1995Jul10      &    1415 &   309 &  6    \\
VLA-BnC   &    17.6            & 2000Mar13      &    1425 &   308 &  1    \\
VLBA      &  0.0048            & 1997Jan10$-$12 &    2320 &   270 &  3    \\
VLA-B     &     2.0            & 1994Aug06      &    4710 &   408 &  6    \\
VLA-C     &     4.4            & 1980Jul13,14   &    4885 &   337 &  2    \\
VLBI      &  0.0014            & 1993Jun9$-$16  &    4992 &   461 &  7    \\
VLA-A     &     0.2            & 1990Feb19$-$23 &    8400 &   673 &  5    \\
VLBA      &  0.0012            & 1997Jan10$-$12 &    8550 &   290 &  3    \\
VLA-CnD   &     8.0            & 1997Sep-Oct    &    8460 &   458 &  4    \\
VLA-C     &     1.4            & 1980Jul13,14   &   15035 &   465 &  2    \\
\hline
\end{tabular}
1: Present paper; 2: Saikia et al. 1984; 3: Fey et al. 2000;
4: Dennett-Thorpe et al. 2000; 5: Patnaik et al. 1992; 6: Taylor et al. 1996;
7: Henstock et al. 1995.
\end{table}

\section{Concluding remarks}
Multifrequency GMRT and VLA observations      
of a sample of seventeen largely GRSs 
have either helped clarify the radio structures or provided new
information at a different frequency.  The broad line
radio galaxy, J0313+413, has an asymmetric, curved radio jet and a variable radio
core, consistent with a moderate angle of inclination to the line of sight, while J1604+375
shows evidence of a precessing twin-jet structure. 
We identify four candidate steep spectrum radio cores, which may be a sign of recurrent activity.
J0139+399, J1235+213 and
J1919+517 appear to have $\alpha_{\rm core}\lapp-$0.8, while J1637+417 is a more marginal case with
$\alpha_{\rm core}\sim-$0.5. The spectra of these candidate SSCs need to be determined over a larger
frequency range from higher resolution observations. If confirmed, this would indicate a higher incidence of SSCs
among GRSs, compared with sources of smaller dimensions, which is possibly related to the time scales of recurrent activity.
From the  structure and integrated spectra of the sources, we suggest that the diffuse lobes of emission
in J0139+399 and J0200+408 may be due to an earlier cycle of nuclear activity.
We find that inverse-Compton losses with the cosmic microwave background radiation dominate
over synchrotron radiative losses in the lobes of all the sources, consistent 
with earlier studies. We also show that the prominence of the bridge emission 
decreases with redshift, possibly due to inverse-Compton losses. This would affect
the appearance and identification of GRSs at large redshifts.

\section*{Acknowledgments}
We thank an anonymous referee for a detailed report and many useful suggestions, and 
Ian Browne and Graham Smith for several useful comments on the manuscript, all of which
have significantly improved the paper.
The Giant Metrewave Radio Telescope is a national facility operated by the National Centre 
for Radio Astrophysics of the Tata Institute of Fundamental Research.
The National Radio Astronomy Observatory  is a
facility of the National Science Foundation operated under co-operative
agreement by Associated Universities Inc. 
This research has made use of the NASA/IPAC extragalactic database (NED)
which is operated by the Jet Propulsion Laboratory, Caltech, under contract
with the National Aeronautics and Space Administration.
DJS thanks the Director, Jodrell Bank Observatory for hospitality, where
this work was completed.

{}

\end{document}